\documentclass[6pt,draftclsnofoot,onecolumn]{IEEEtran}

\usepackage{amsmath}
\usepackage{amssymb}
\usepackage{amsfonts}
\usepackage{graphics}
\usepackage{pgf}
\usepackage{bbm}
\usepackage{lipsum}

\usepackage{enumitem}

\usepackage{tikz}
\usetikzlibrary{positioning,shapes}
\usepackage{subfigure}

\newtheorem{theorem}{Theorem}

\newtheorem{definition}[theorem]{Definition}
\newtheorem{example}[theorem]{Example}

\newtheorem{lemma}[theorem]{Lemma}
\newtheorem{remark}[theorem]{Remark}
\newenvironment{proof}[1][Proof]{\noindent\textbf{#1.} }{\ \rule{0.5em}{0.5em}}

\def\tr{\textrm{Tr}}
\def\id{\textrm{id}}
\def\pr{\textrm{Pr}}
\def\mod{\textrm{mod}}
\def\1{\mathbbm{1}}
\def\ox{\otimes}
\def\be{\begin{equation}}
\def\ee{\end{equation}}
\def\ba{\begin{eqnarray}}
\def\ea{\end{eqnarray}}

\newcommand{\?}{\stackrel{?}{=}}

\newcommand{\ket}[1]{|#1 \rangle}
\newcommand{\bra}[1]{\langle #1|}
\newcommand{\proj}[1]{|#1 \rangle \! \langle #1 |}
\newcommand{\SWAP}{\operatorname{SWAP}}

\newcommand{\conf}{\operatorname{C_{conf}}}
\newcommand{\conften}{\operatorname{C_{conf\otimes}}}

\let\originalleft\left
\let\originalright\right
\renewcommand{\left}{\mathopen{}\mathclose\bgroup\originalleft}
\renewcommand{\right}{\aftergroup\egroup\originalright}

\usepackage{url}

\usepackage{color}

\begin{document}

\title{Classical capacities of quantum channels with environment assistance}

\author{Siddharth Karumanchi,
\thanks{Siddharth Karumanchi is with the Faculty of Informatics, Masaryk University, Botanick\'{a} 68a, Brno, Czech Republic. This work was done when he
 was with the
School of Science and Technology,
University of Camerino,
Via M. delle Carceri 9, I-62032 Camerino, Italy and
INFN--Sezione Perugia,
Via A. Pascoli, I-06123 Perugia, Italy. Email: siddharth.karumanchi@unicam.it} Stefano Mancini, \thanks{Stefano Mancini is with the
School of Science and Technology,
University of Camerino,
Via M. delle Carceri 9, I-62032 Camerino, Italy and
INFN--Sezione Perugia,
Via A. Pascoli, I-06123 Perugia, Italy. Email: stefano.mancini@unicam.it}
Andreas Winter, \thanks{Andreas Winter is with ICREA and
F\'{\i}sica Te\`{o}rica: Informaci\'{o} i Fen\`{o}mens Qu\`{a}ntics,
Universitat Aut\`{o}noma de Barcelona,
ES-08193 Bellaterra (Barcelona), Spain. Email: andreas.winter@uab.cat}
and Dong Yang\thanks{Dong Yang is with
F\'{\i}sica Te\`{o}rica: Informaci\'{o} i Fen\`{o}mens Qu\`{a}ntics,
Universitat Aut\`{o}noma de Barcelona,
ES-08193 Bellaterra (Barcelona), Spain and
Laboratory for Quantum Information,
China Jiliang University,
Hangzhou, Zhejiang 310018, China. Email: dyang@cjlu.edu.cn}
}



\maketitle

\begin{abstract}
A quantum channel physically is a unitary interaction between the
 information carrying system and an environment, which is initialized
 in a pure state before the interaction. Conventionally, this state, as
 also the parameters of the interaction, is assumed to be fixed and
 known to the sender and receiver. Here, following the model 
 introduced by us earlier [Karumanchi \emph{et al.}, arXiv[quant-ph]:1407.8160], we consider
 a benevolent third party, i.e. a helper, controlling the environment
 state, and how the helper's presence changes the communication
 game. In particular, we define and study the classical capacity of a
 unitary interaction with helper, indeed two variants, one where the
 helper can only prepare separable states across many channel
 uses, and one without this restriction. Furthermore, the two even
 more powerful scenarios of pre-shared entanglement between
 helper and receiver, and of classical communication between sender
 and helper (making them conferencing encoders) are considered. 
\end{abstract}


\IEEEpeerreviewmaketitle

\section{Introduction}
\label{sec:intro}
The noise in quantum communication is modelled by a quantum channel, which is a completely positive and trace preserving (CPTP) map on the set of states (density operators) of a system devoted to carry information. Note that this view contains classical channels as a special case (cf.~\cite{Wilde11}). Every quantum channel can be viewed as a unitary interaction between the information carrying system and an environment, where the latter is customarily considered not under control. Therefore, the initial environment state together with the unitary defines the channel when the final environment is traced out.
In this standard picture, the initial environment state is simply fixed, and the environment output is completely lost. However, the possibility of an \emph{active helper}, one that reads information from the channel environment and communicates it to the channel receiver, is an interesting one that has been considered before with some success~\cite{GW03, GW04, HK05,SVW05,Winter07}. In the present paper, instead, we shall be concerned with a benevolent party (a helper, hence called Helen) setting the initial environment state in order to assist sender and receiver of the channel to communicate.
 We considered transmission of quantum information in this model in our earlier work~\cite{KMWY14}. Here we look at classical communication: as in~\cite{KMWY14}, we have a model of \emph{passive environment assistance}, where Helen simply sets an initial state of the environment as part of the code, once and for all; likewise, we motivated to consider \emph{passive environment assistance with entanglement} between Helen and the receiver Bob of the channel output. Because classical information, unlike quantum information, can be freely shared, here we can then contrast these passive models with one where Helen's state can also depend on the message to be sent; we call it \emph{conferencing encoders}, allowing local operations and classical communication (LOCC) between the sender Alice and Helen.

The structure of the paper is as follows: Section II introduces the  notation and provides the details of the proposed models.
 Section III contains the coding theorems of passive environment assisted capacities.  There, after making general observations, we also provide examples of super-additivity of capacities. Then, in Section IV we go on to study the entanglement environment assisted capacities. In the following Section V we make general observations about the conferencing encoders model, and go on to show that for a unitary operator the classical capacity with conferencing encoders is non-zero.
 Finally, the appendix give details of the parametrization for two-qubit unitaries (Appendix~\ref{Upara}).

\section{Notation and models}

Let $A$, $E$, $B$, $F$, etc.~be finite dimensional Hilbert spaces, $\mathcal{L}(X)$ denote the space of linear operators on the 
Hilbert space $X$ and $|X|$ denote the dimension of the Hilbert space. \\
We denote the identity operator in $\mathcal{L}(X)$ as $\1 ^{X}$ and the ideal map, $\id : \mathcal{L}(X) \rightarrow \mathcal{L}(X)$ is denoted by $\id^X$. For any linear operator $\Lambda : A \rightarrow B$ we will use the \emph{trace norm} defined as
\begin{equation}
\label{eq:tracenorm}
\left\|\Lambda \right\|_1 = \tr\sqrt{\Lambda^{\dag} \Lambda} = \tr\vert \Lambda \vert,
\end{equation}
and the \emph{operator norm} defined as
\begin{equation}
\label{eq:opnorm}
\left\|\Lambda \right\|_{\rm op} = \sup \left\{ \left\|\Lambda a \right\| : a \in A, \left\| a \right\| = 1  \right\}.
\end{equation}
For any super-operator $\mathcal{N} : \mathcal{L}(A) \rightarrow \mathcal{L}(B)$ the \emph{induced trace norm} is defined as
\begin{equation}
\label{eq:inducedtracenorm}
\left\| \mathcal{N} \right\|_{1\rightarrow 1} = \max \left\{ \left\| \mathcal{N}(\rho) \right\|_{1} ; \rho \in \mathcal{L}( A) ; \left\| \rho \right\|_1 = 1  \right\}.
\end{equation}
Furthermore, for any super-operator $\mathcal{N} : \mathcal{L}(A) \rightarrow \mathcal{L}(B)$ we will make use of the \emph{diamond norm} defined as
\begin{equation}
\label{eq:diamondnorm}
\left\| \mathcal{N} \right\|_\diamond = \left\| \id^A \ox \mathcal{N} \right\|_{1\rightarrow 1} = \max \left\{ \left\| \id^{A} \ox \mathcal{N}(\rho) \right\|_{1} ; \rho \in \mathcal{L}(A \ox A) ; \left\| \rho \right\|_1 = 1  \right\}.
\end{equation}
Note that the maximum in this definition is attained on a rank-one operator $\rho = \ket{\psi} \bra{\varphi}$, with unit vectors $\ket{\psi}$ and $\ket{\varphi}$. If $\mathcal{N}$ is Hermitian-preserving, then the maximum is indeed attained on a pure state $\rho = \proj{\psi}$. For any super-operator $\mathcal{N} : \mathcal{L}(A) \rightarrow \mathcal{L}(B)$ the diamond norm and the induced trace norm are related as follows:
\begin{equation}
\label{eq:diamondtracenorm}
\left\| \mathcal{N} \right\|_{1\rightarrow 1} \leq \left\| \mathcal{N} \right\|_\diamond \leq \left\| \mathcal{N} \right\|_{1\rightarrow 1} \min(\vert A\vert, \vert B\vert).
\end{equation}
For a density operator $\alpha^A$ the \emph{von Neumann entropy} is defined as 
\begin{equation}
\label{eq:entropy}
S(A)_{\alpha} = S(\alpha) := - \tr \alpha \log \alpha.
\end{equation}
For two density operators $\alpha$ and $\beta$ the \emph{quantum relative entropy} of $\alpha$ with respect to $\beta$ is defined as
\begin{equation}
\label{eq:relativeentropy}
D(\alpha || \beta) := \tr \alpha (\log \alpha - \log \beta).
\end{equation}
Furthermore, for any density operator $\rho^{AB}$ on a bipartite system, the \emph{quantum mutual information} is defined as
\begin{equation}
\label{eq:mutualinfo}
I\bigl( A : B \bigr)_{\rho} := S(A)_{\rho} + S(B)_{\rho} - S(AB)_{\rho}.
\end{equation}

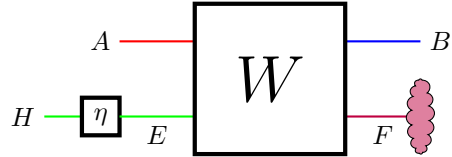
\begin{figure}[ht]
\begin{center}
\begin{tikzpicture}[scale=0.5]
\draw [thick, green](0,0) -- (1,0);
\draw[ultra thick] (1,-0.5) rectangle (2,0.5) node[midway]{$\eta$};
\draw[thick, green] (2,0) -- (4,0); \draw[thick, red] (2,2) -- (4,2); \draw[thick, blue] (8,2) -- (10,2);
\draw[thick,purple] (8,0) -- (10,0);
\draw[ultra thick] (4,-1) rectangle (8,3);
\node[cloud, cloud puffs=15.7, cloud ignores aspect, minimum width=0.3cm, minimum height=1cm, align=center, draw, fill=purple!50] (cloud) at (10,0){} ; 
\node[left] at (0,0){$H$}; \node[right] at (10,2){$B$};
\node[left] at (2,2){$A$}; \node[below] at (3,0){$E$};
\node[below] at (9,0){$F$};
\draw (6,1) node[font = \fontsize{40}{42}\sffamily\bfseries]{$W$};
\end{tikzpicture}
\end{center}
  \caption{Diagrammatic view of the three parties, Alice ($A$), Helen ($H$) and Bob ($B$), involved in the communication with a third party ($H$) controlling the environment input system  whose aim is to enhance the communication between Alice and Bob. As Helen has no further role to play after setting the initial environment state, its assistance is of passive nature. The inaccessible output-environment system is labelled as $F$.}
  \label{fig:modelpassive}
\end{figure}

We have three presumably inequivalent models of classical communication, depending on role of the helper. In the first model under consideration, we assume that  Helen sets the initial state of the environment to enhance the classical communication from Alice to Bob as depicted in Fig. \ref{fig:modelpassive}.  Since Helen has no role in the protocol after setting the initial environment state, this model is thus referred to as \emph{passive environment-assisted model}.

We assume that there are no quantum correlations between Alice's and Helen inputs. 
Consider a unitary or more generally an isometry $W:A\ox E \longrightarrow B\ox F$, which
defines the channel (CPTP map) 
$\mathcal{N}:\mathcal{L}(A\ox E) \rightarrow \mathcal{L}(B)$,
whose action on the input state $\rho$ on $A\ox E$ is
\begin{equation}
\label{eq:church}
  \mathcal{N}^{AE \rightarrow B}(\rho) = {\tr}_F W \rho W^\dag.
\end{equation}
Then, an effective channel 
$\mathcal{N}_\eta:\mathcal{L}(A) \rightarrow \mathcal{L}(B)$ is established between Alice and Bob
once the initial state $\eta$ on $E$ is set:
\begin{equation}
\label{eq:effective-channel}
  \mathcal{N}_\eta^{A\rightarrow B}(\rho) := \mathcal{N}^{AE\rightarrow B}(\rho\ox\eta).
\end{equation}

For a given CPTP map $\mathcal{N} : \mathcal{L}(A) \rightarrow \mathcal{L}(B)$, we can consider the Stinespring isometry $V : A \longrightarrow B \ox F$. This is a special case of the above model where the initial environment $E$ is \emph{one-dimensional}, i.e. Helen has no choice of the initial environment state.
The classical capacity for the above case is given by the Holevo-Schumacher-Westmoreland theorem~\cite{Holevo98, SW97} (cf.~\cite{Wilde11}), 
\begin{equation}
\begin{split}
    \label{eq:CC}
    C(\mathcal{N}) &= \sup_{n} \max_{\{ p_x , \rho^{A^n}_x \}}
                   \frac{1}{n} I\bigl( X : B^n \bigr)_{\sigma},
  \end{split} 
  \end{equation}
  where the quantum mutual information is evaluated with respect to the state 
 \begin{equation}
  \label{eq:cqstate}
  \sigma = \sum\limits_{x}p_{x}\proj{x} \ox \mathcal{N}^{\ox n}(\rho^{A^n}_{x}).
 \end{equation}
   The maximization is over the ensembles $\{ p_x , \rho^{A^n}_x \}$ where the states  $\rho^{A^n}_x$ are input across $A^n$ . Here $\{ \ket{x} \}$ are the orthonormal basis of the classical reference system $X$. It is known that
  the supremum over $n$ (the ``regularization'') is necessary~\cite{Hastings09},
  except for some special channels~\cite{AM09}.
  
  When the encoding by the sender is restricted to separable states $\rho_x^{A^n}$, i.e. convex combinations of tensor products $\rho_x^{A^n} = \rho_{1x}^{A_1} \ox \ldots \ox \rho_{nx}^{A_n}  $, the classical communication capacity admits a single-letter  characterization, given by the so-called \emph{Holevo information} of quantum channel,
  \begin{equation}
  \begin{split}
    \label{eq:Holevo}
    \chi(\mathcal{N}) &= \max_{\{p_x, \rho_{x}^A \}}
                    I\bigl( X : B \bigr)_{\sigma},
  \end{split}
\end{equation}
where the quantum mutual information is evaluated with respect to the state $\sigma := \sum\limits_{x} p_x \proj{x} \ox \mathcal{N}(\rho_x^A)$. 

The ensemble that achieves the maximum in Eq.~(\ref{eq:Holevo}), say $\{ p^{*}_x, \phi^{*}_x \}$, is called the \emph{optimal ensemble}.
   Let $\phi_{avg}^{*} = \sum\limits_{x} p^{*}_{x}\phi^{*}_{x}$ be the average of the optimal ensemble. From~\cite{SW01} we know the existence of such an ensemble that achieves the Holevo information, and has the following relative entropic formulation,
\begin{equation}
\label{eq:Holevorelative1}
\chi(\mathcal{N}) = \sum\limits_{x} p^{*}_x D(\mathcal{N}(\phi^{*}_x) || \mathcal{N}(\phi_{avg}^{*})).
\end{equation}
For any $\rho^{A}$, we have the following inequality,
\begin{equation}
\label{eq:Holevorelative2}
\chi(\mathcal{N}) \geq  D(\mathcal{N}(\rho^{A}) || \mathcal{N}(\phi_{avg}^{*})),
\end{equation}
with the equality holding for any member of the optimal ensemble. Also note that for a given channel $\mathcal{N}$, though we can have many optimal ensembles that achieve the Holevo information, the average of the optimal ensemble  is unique~\cite{Cortese03}.
An important class of channels which admit single-letter characterization of classical capacity i.e.~$C(\mathcal{N}) = \chi(\mathcal{N})$, are the 
 entanglement-breaking channels~\cite{HSR03}. Actually, for any two  entanglement-breaking channels, $\mathcal{N}_1$ and $\mathcal{N}_2$, we have the following additivity property :
\begin{equation}
\label{eq:additivityeb}
\chi(\mathcal{N}_1 \ox \mathcal{N}_2) = \chi(\mathcal{N}_1) + \chi(\mathcal{N}_2).
\end{equation}

A variant of the passive environment-assisted model where Helen has pre-shared entanglement with Bob, thus referred to as \emph{entanglement-environment-assisted model,} can also be considered.
In such a case we can extend 
the notation of $\mathcal{N}_\eta =  \mathcal{N}(\bullet \ox \eta)$
and let, for a state $\kappa$ on $EK$,
\begin{equation}
\label{eq:entassistmap}
  \mathcal{N}_\kappa^{A \rightarrow BK}(\rho) 
         := (\mathcal{N}^{AE \rightarrow B}\ox{\id}^K)(\rho^A\ox\kappa^{EK}).
\end{equation}

\begin{figure}[ht]
\begin{center}
\begin{tikzpicture}[scale=0.5]
\draw [thick, green](0,0) -- (4,0);
\draw[<->, line width = 2] (2,0) -- (2,2);
 \draw[thick, red] (0,2) -- (4,2); \draw[thick, blue] (8,2) -- (10,2);
 \draw[thick, purple] (8,0) -- (10,0);
\draw[ultra thick] (4,-1) rectangle (8,3);
\node[cloud, cloud puffs=15.7, cloud ignores aspect, minimum width=0.3cm, minimum height=1cm, align=center, draw, fill=purple!50] (cloud) at (10,0){} ; 
\node[left] at (0,0){$H$}; \node[right] at (10,2){$B$};
\node[above] at (3,2){$A$}; \node[below] at (3,0){$E$};
\node[below] at (9,0){$F$};
\draw (6,1) node[font = \fontsize{40}{42}\sffamily\bfseries]{$W$};

\end{tikzpicture}
\end{center}
  \caption{Diagrammatic view of the three parties
   Alice ($A$), Helen ($H$) and Bob ($B$), involved in the communication, with  the party ($H$) 
   controlling the environment input system and the sender ($A$) that can freely communicate classically. The inaccessible output-environment system is labelled as $F$. }
  \label{fig:modelconf}
\end{figure}
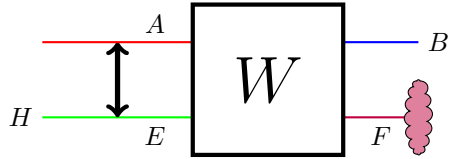

A new model called \emph{conferencing helper} is introduced, where we contemplate the possibility of local operations and classical communication (LOCC) between Alice and Helen, thus allowing Helen to play an active role in the encoding process (in contrast to the previous models discussed above),  see Fig. \ref{fig:modelconf}. 
 Of course the possibility that Alice and Helen share entanglement has to be excluded otherwise we recover the situation of a quantum channel determined by Alice and Helen input system and Bob output obtained by tracing away part of the system. For a given unitary or more generally an isometry $W:A\ox E \longrightarrow B\ox F$, which
defines the channel 
$\mathcal{N}:\mathcal{L}(A\ox E) \rightarrow \mathcal{L}(B)$,
whose action on a input state is 
\begin{equation}
\label{eq:confmap}
\mathcal{N}^{AE\rightarrow B}(\rho_i \ox\eta_i) = {\tr}_F W (\rho_i \ox\eta_i )W^\dag,
\end{equation}
the state $\eta_i$ can be adjusted according to the classical message with the proviso that
the global input state of the systems $A$ and $E$ is separable. 

 Furthermore, we would mention that 
throughout the paper $\log$ is intended as logarithm on base $2$ and $\ln$ the natural logarithm. The binary entropy is denoted by
\begin{equation}
\label{eq:binaryentropy}
H_{2}(p) = -p \log (p) - (1-p) \log (1-p).
\end{equation} 

The cyclic shift operator $X(x)$ and the phase operator $Z(z)$ acting on the computational basis 
$\{\vert j \rangle\}_{0,1,2, ... ,d-1}$ of a $d$-dimensional Hilbert space are defined in the following way
\begin{equation}
\label{eq:cyclicphaseoperators}
\begin{split}
X(x) \vert j \rangle  &=  \ket{(x+j) \mod \, d}, \quad x=0,1,2, ... ,d-1, \\
Z(z) \vert j \rangle  &=  \omega^{zj}\vert j \rangle, \quad z=0,1,2, ... ,d-1.
\end{split}\end{equation}
Here the complex number $\omega = \exp (\frac{ 2 \pi i }{d})$ is a primitive $d$-th root of unity and $i$ denotes the imaginary unit. Given the operators in Eq.~(\ref{eq:cyclicphaseoperators}), for each pair $(x,z)$, we can identify the \emph{discrete Weyl operator} $W(x,z) \in \rm U(d)$ defined as
\begin{equation}
\label{eq:weyloperator}
W(x,z) := X(x)Z(z).
\end{equation}


\section{Passive Environment Assisted Capacities}\label{sec:passive}

In this Section we define the \emph{passive environment-assisted} model rigorously and provide different notions of assisted codes, depending on the capabilities of Helen (whether she can input arbitrary entangled states across different instances of isometry or whether she is restricted to separable states). Furthermore, a capacity when Helen is restricted to separable states and Alice's encoding is restricted to product state across $n$ instances of the channel is also considered.


\subsection{Model for transmission of classical information}\label{sec:model}

\begin{figure}[ht]
\begin{center}
\begin{tikzpicture}[scale=0.3]
\draw [ultra thick] (2,-5) rectangle (7,5) ;
\draw [ultra thick] (15,-5) rectangle (20,5) ;
\draw [ultra thick] (10,3) rectangle (12,5) node[midway]{$\mathcal{N}$};
\draw [ultra thick] (10,0) rectangle (12,2) node[midway]{$\mathcal{N}$};
\draw [ultra thick] (10,-5) rectangle (12,-3) node[midway]{$\mathcal{N}$};
\draw [ultra thick, red] (0,0)   -- (2,0);
\draw [ultra thick, blue] (20,0)   -- (22,0);
\draw [thick, red] (7,4.5) -- (10,4.5);
\draw [thick, red] (7,1.5) -- (10,1.5);
\draw [thick, red] (7,-3.5) -- (10,-3.5);
\draw [thick,blue] (12,4) -- (15,4);
\draw [thick, blue] (12,1) -- (15,1);
\draw [thick, blue] (12,-4) -- (15,-4);
\draw [thick, green] (8,-7) -- (8,3.5) -- (10,3.5);
\draw [thick, green] (8,-7) -- (8.2,0.5) -- (10,0.5);
\draw [thick, green] (8,-7) -- (8.4,-4.5) -- (10,-4.5);

\node[left] at (0,0){$M$};
\node[right] at (22,0){$\tilde{M}$};
\node[above] at (7.5,4.5){$A$};\node[above] at (7.5,1.5){$A$};\node[above] at (7.5,-3.5){$A$};
\node[below] at (9.3,3.5){$E$};\node[below] at (9.3,0.5){$E$};\node[below] at (9.3,-4.5){$E$};\node[above] at (14,4){$B$};\node[above] at (14,1){$B$};\node[above] at (14,-4){$B$};
\node[below] at (8,-7){$\eta$};
\draw (4.5,0) node[font = \fontsize{40}{42}\sffamily\bfseries]{$\mathcal{E}$};
\draw (17.5,0) node[font = \fontsize{40}{42}\sffamily\bfseries]{$\mathcal{D}$};
\draw [thick,dotted] (7.5,-0.5) -- (7.5,-1.5);
\draw[thick,dotted] (9.3,-1.5) -- (9.3,-2.5);
\draw[thick,dotted] (11,-0.9) -- (11,-1.9);
\draw[thick,dotted] (14,-1) -- (14,-2);
\end{tikzpicture}
\end{center}
\caption{Schematic of a general protocol to transmit classical information with passive assistance from the environment; $\mathcal{E}$ 
    and $\mathcal{D}$ are the encoding and decoding maps respectively. 
    The initial state of the environment is $\eta$.}
\label{fig:passiveinfotask}
\end{figure}
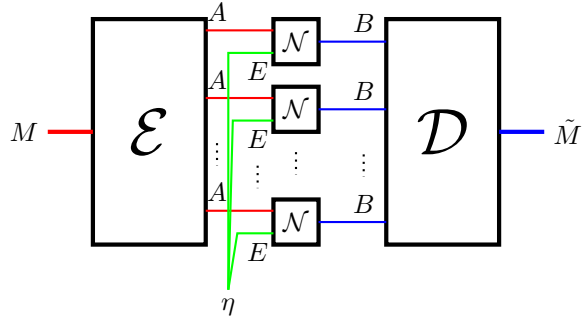

 By referring to Fig. \ref{fig:passiveinfotask},
let Alice selects some classical message $m$ from the set of messages $\{1,2, ..., \vert M  \vert \}$ to communicate to Bob. Let $\mathsf{M}$ denote the random variable corresponding to Alice's choice of message and $M$ corresponds to the associated Hilbert space with the orthonormal basis $\{ \ket{m} \}$. 
An encoding CPTP map 
$\mathcal{E}: M \rightarrow \mathcal{L}(A^n)$ can be realised by preparing states $\{ \alpha_m \}$ to be input across $A^n$ of $n$ instances of the channel. A decoding CPTP map $\mathcal{D}:\mathcal{L}(B^n) \rightarrow \tilde{M}$ can be realised by a positive operator-valued measure (POVM) $\{ \Lambda_{m} \}$. Here $\tilde{M}$ is the Hilbert space associated to the random variable $\tilde{\mathsf{M}}$ for Bob's estimate of the message sent by Alice. The probability of error for a particular message $m$ is 
\begin{equation}
\label{eq:proboferror}
P_e(m)= 1 - \tr\left[\Lambda_m
\mathcal{N}^{\ox n}(\alpha_{m}^{A^n} \ox \eta^{E^n})\right].
\end{equation}

\bigskip

\begin{definition}
 A \emph{passive environment-assisted classical code} of block length $n$ is a family of triples
  $\{\alpha_{m}^{A^n},\eta^{E^n},\Lambda_{m}\}$ with the error probability $\overline{P_e} :=\frac{1}{|M|}\sum_m P_e(m)$  and the rate $\frac{1}{n}\log |M|$.
  A rate $R$ is achievable if there is a sequence of codes over their block length $n$ with $\overline{P_e}$ converging to $0$ and rate converging to $R$. The passive environment-assisted classical capacity of $W$, denoted by $C_{H}(W)$ or equivalently $C_{H}(\mathcal{N})$, is the maximum achievable rate. If the helper is restricted to fully separable states $\eta^{E^n}$, i.e.~convex
  combinations of tensor products $\eta^{E^n} = \eta_1^{E_1} \ox \cdots \ox \eta_n^{E_n}$,
  the largest achievable rate is denoted $C_{H\ox}(W) = C_{H\ox}(\mathcal{N})$.
\end{definition}

As the error probability is linear in the environment state $\eta$, without loss of
generality $\eta$ may be assumed to be pure, for both unrestricted and
separable helper. We shall assume this from now on, without necessarily specifying it each time.

\bigskip

\begin{theorem}
  \label{CH+CHtens}
  For an isometry $W:AE \longrightarrow BF$, the passive environment-assisted
  classical capacity is given by 
  \begin{equation}\begin{split}
    \label{eq:CH}
    C_{H}(W) &= \sup_{n} \max_{\eta^{(n)}} \frac1n C(\mathcal{N}^{\ox n}_{\eta^{(n)}}) \\
             &= \sup_{n} \max_{ \{p(x),\alpha_{x}^{A^n}\},\eta^{E^n}}
                   \frac{1}{n} I\bigl( X : B^n \bigr)_{\sigma},
  \end{split}\end{equation}
  where the mutual information is evaluated with respect to the state 
  \begin{equation}
  \sigma = \sum\limits_{x} p(x) \proj{x} \ox \mathcal{N}^{\ox n}_{\eta^{E^n}}(\alpha^{A^n}_{x})
  \end{equation}
  and  the maximization is over the ensemble $\{ p(x) , \alpha^{A^n}_{x} \}$
 and pure
  environment input states $\eta^{(n)}$ on $E^n$.
  
  Similarly, the capacity with separable helper is given by the formula,
  \begin{equation}\begin{split}
    \label{eq:CHtens}
    C_{H\ox}(W) &= \sup_{n} \max_{\eta^{(n)}=\eta_1\ox\cdots\ox\eta_n}
                     \frac1n C(\mathcal{N}_{\eta_1}\ox\cdots\ox\mathcal{N}_{\eta_n})   \\
                &= \sup_{n} \max_{\{p(x),\alpha_{x}^{A^n}\},\eta^{E^n}}
                   \frac{1}{n} I\bigl( X : B^n \bigr)_{\sigma},
  \end{split}\end{equation}
  where the maximum is only over (pure) product states, 
  i.e.~$\eta^{(n)} = \eta_1 \ox \cdots \ox \eta_n$.
  
  As a consequence, $C_H(W) = \lim_{n\rightarrow\infty} \frac1n C_{H\ox}(W^{\ox n})$.
\end{theorem}

\medskip

\begin{proof}
The direct part (the ``$\geq$'' inequality), follows directly from the 
  HSW theorem~\cite{Holevo98,SW97}, applied to the channel
  $(\mathcal{N}^{\ox n})_{\eta^{(n)}}$; to be precise asymptotically many
  copies of this block-channel, so that the i.i.d.~arguments hold true (cf.~\cite{Wilde11}).
  
   For the converse  part (the ``$\leq$'' inequality), consider a code of block length $n$ with  error probability $\overline{P_e}$. The state after encoding operation and action of the channel  is given by  
\begin{equation}
\Phi^{M B^n} = \frac{1}{|M|} \sum\limits_{m}p(m) \proj{m} \ox \mathcal{N}^{\ox n}(\alpha^{A^n}_{m} \ox \eta^{E^n}) ,
\end{equation}
and the state after decoding operation is given by
\begin{equation}
\omega^{M \tilde{M}} = \1^{M} \ox \mathcal{D}(\Phi^{M B^n}).
\end{equation}
Then we have:
\begin{equation}
\begin{split}
 nR &= H(M)_{\omega}, \\
           &= I(M : \tilde{M})_{\omega} + H(M \vert\tilde{M})_{\omega}, \\
     &\leq I(M : \tilde{M})_{\omega} + H(\overline{P_e}) + nR\overline{P_e} ,\\
     &\leq I(M : B^n)_{\Phi} + n\epsilon.
\end{split}
\end{equation}
The first inequality follows from  the application of Fano's inequality and the second one follows from  the data processing inequality, where $\epsilon  = \frac{1}{n} + R \overline{P_e}$. Setting $M = X$ we have
\begin{equation}
R \leq \frac{1}{n} I(X : B^n) + \epsilon .
\end{equation}
 As $n \rightarrow \infty$ and $\overline{P_e} \rightarrow 0$, the upper bound on the rate
  follows -- depending on $C_H$ or $C_{H\ox}$, without or with restrictions
  on $\eta^{(n)}$.
\end{proof}

\bigskip

\begin{remark}
The channel whose inputs are Alice and Helen and outputs Bob can be viewed as  a quantum version of multiple access channels (MAC) with two senders and one receiver which was studied in~\cite{W01,YHD08}. In such a model both Alice and Helen try to communicate their individual independent messages to Bob. The rates  $R_A, R_H$ at which Alice and Helen can respectively communicate with Bob gives the capacity region $(R_A, R_H)$. If this capacity region is known, then the passive environment assisted capacity is given by $\max \{ R : (R,0) \in \text{capacity region} \}$.  Whenever single letter characterization  for a MAC is available this might be helpful in the evaluation of environment assisted capacities, but in general when the regularization is required this view may not help.
\end{remark}

\bigskip

 For separable helper, and when in addition Alice's encoding is restricted to product input states, i.e. 
 \begin{equation}
 \alpha_m = \alpha_{1m} \ox \alpha_{2m} \ldots \ox \alpha_{nm},
 \end{equation}
  across $A^n$, we have the following:
\begin{equation}
\max_{\{p(x), \alpha_{x}^{A^n} \}, \eta_1\ox\cdots\ox\eta_n}
                    I\bigl( X : B^n \bigr)_{\sigma} = \sum\limits_i \max_{\{p(x_i), \alpha_{x_i}^{A} \} , \eta_i} I \bigl( X_i : B \bigr)_{\sigma_i}.
\end{equation} 
  Then, from Eq.~(\ref{eq:CHtens}), we have the product state capacity with separable helper given by
  \begin{equation}\begin{split}
    \label{eq:CHprodtens}
    \scalebox{1.5}{$\chi$}_{H\ox}(W) &=\max_{\{ p(x) , \rho_x \},\eta^{E}}
                   I\bigl( X : B \bigr)_{\sigma},
  \end{split}\end{equation}
  where the mutual information is evaluated with respect to the state $\sigma := \sum\limits_{x} p(x) \proj{x} \ox \mathcal{N_{\eta}}(\rho_x)$ and the maximization is over the ensemble $\{ p(x) , \rho_x \}$ and the state $\eta^E$.

  For any $\eta^{E}$ and for all $\rho ^A$ it is
 \begin{equation}\label{eq:relativechihox}
 D(\mathcal{N_{\eta}}^{A \rightarrow B}(\rho ^A) || \omega_{\eta}^{B})  \leq  \scalebox{1.5}{$\chi$}_{H\ox}(W),
 \end{equation}
 where $\omega_{\eta}^{B} := \mathcal{N}_{\eta}^{A\rightarrow B}(\rho_{avg,\eta} ^A)$ and $\rho_{avg,\eta}^{A}$ is the average of the optimal ensemble that achieves the Holevo information for $\mathcal{N_{\eta}}$.

We can further contemplate the scenario where the roles of $A$ and  $E$ are exchanged, i.e. Alice tries to set an initial state in $A$, thereby establishing a channel between $E$ and $B$. Then, the quantities of interest are $\scalebox{1.5}{$\chi$} ^{A}_{H \ox}$,  $\scalebox{1.5}{$\chi$}^{H}_{A\ox}$, the product state capacities with a separable helper when the sender is Alice and Helen respectively. They are given by the following formulae \footnote{ When we omit the 
superscripts, it is implicitly understood that Alice is the sender and Helen sets an initial state in $E$.}
 \begin{equation}\begin{split}
    \label{eq:CHprodtensAE}
    \scalebox{1.5}{$\chi$}^{A}_{H\ox}(W) &=\max_{\{ p(x) , \alpha_x^{A} \},\eta^{E}}
                   I\bigl( X : B \bigr)_{\sigma}, \\
  \scalebox{1.5}{$\chi$}^{H}_{A\ox}(W) &=\max_{\{ p(x) , \eta_x^{E} \},\alpha^{A}}
                   I\bigl( X : B \bigr)_{\mu},
  \end{split}\end{equation}
 where the mutual information for the former case is evaluated with respect to the state $\sigma := \sum\limits_{x} p(x) \proj{x} \ox \mathcal{N}_{\eta}(\alpha_x)$ and the maximization is over the ensemble $\{ p(x) , \alpha_x \}$ and the state $\eta^E$. In the latter case, when Helen is the sender, the mutual information is evaluated with respect to the state $\mu := \sum\limits_{x} p(x) \proj{x} \ox \mathcal{M}_{\alpha}(\eta_x)$ and the maximization is over the ensemble $\{ p(x) , \eta_x \}$ and the state $\alpha^A$. Here, the effective channels $\mathcal{N}_{\eta} : \mathcal{L}(A) \rightarrow \mathcal{L}(B)$ and $\mathcal{M}_{\alpha} : \mathcal{L}(E) \rightarrow \mathcal{L}(B)$ are respectively
 \begin{equation}
 \begin{split}
 \mathcal{N}_{\eta}(\rho) = \tr_{F}(W(\rho \ox \eta)W^{\dag}), \\
 \mathcal{M}_{\alpha}(\nu) = \tr_{F}(W(\alpha \ox \nu)W^{\dag}).
 \end{split}
 \end{equation}

 From  Eq.~\ref{eq:CHprodtens} we see that
\begin{equation}
\begin{split}
\label{eq:CHprodtensub}
\scalebox{1.5}{$\chi$}^{A}_{H\ox}(W) &\leq \log \vert B \vert - S_{\rm min}(W), \\
\scalebox{1.5}{$\chi$}^{H}_{A\ox}(W) &\leq \log \vert B \vert - S_{\rm min}(W),
\end{split}
\end{equation}
where
\begin{equation}
\label{eq:minetropy}
S_{\rm min}(W) := \min_{\alpha^{A}, \eta^{E}}S({\tr}_{F}(W(\alpha^{A} \ox \eta^{E})W^{\dag}))
\end{equation}
is the minimum output entropy of the given unitary $W$. 

\bigskip
 
\begin{lemma}[Continuity of the Stinespring dilation][Kretschmann/Schlingemann/Werner~\cite{KSW08}] \label{lemma:continuity}
For any two quantum channels $\mathcal{N}_1, \mathcal{N}_2 : \mathcal{L}(A) \rightarrow \mathcal{L}(B)$ with Stinespring dilations $V_{1}, V_2 : A \longrightarrow B \ox F$, the following holds:
\begin{equation}
 \label{eq:continuity}
   \inf_{U} \left\| (\1^{B} \ox U^F) V_1 - V_2 \right\|_{\rm op}^2 \leq  \left\|  \mathcal{N}_1 - \mathcal{N}_2 \right\|_{\diamond} \leq 2 \inf_{U}\left\|  (\1^{B} \ox U^F) V_1 - V_2   \right\|_{\rm op},
  \end{equation}
  where the infimum is over the unitaries $U : F \longrightarrow F$. 
\end{lemma}

\bigskip

\begin{theorem}
\label{thm:uncertainty}
For any unitary $W :A \ox E \longrightarrow B \ox F$, with $|A|=|E|=|B|=|F|=d$, it holds 
\begin{equation}
\label{eq:uncertainty}
\scalebox{1.5}{$\chi$}_{H\ox}^{A}  + \scalebox{1.5}{$\chi$}_{A\ox}^{H} \geq \frac{1}{2^{13} d^2 \ln 2} \left( \frac{\sqrt{2 + 2( \log d )^2} - \sqrt{2}}{\log d} \right)^8.
\end{equation}
This is a kind of uncertainty relation for $\scalebox{1.5}{$\chi$}_{H\ox}^{A}$ and $\scalebox{1.5}{$\chi$}_{A\ox}^{H}$, saying that not both of them can be arbitrary small.
\end{theorem}  

\medskip

\begin{proof}
Let $\SWAP : A \ox E \longrightarrow B \ox F$ be the \emph{swap operator}, defined by $\SWAP (\ket{\psi}^A \ox \ket{\varphi}^E) := \ket{\varphi}^B \ox \ket{\psi}^F$. 
Let us define a quantum channel which has dilation $\SWAP$ as follows
\begin{equation}
\mathcal{M}_{\sigma}^{A \rightarrow B}(\rho^A) := \tr_F (\SWAP (\rho \ox \sigma ) \SWAP^{\dag}) = \sigma^{B} \tr (\rho^A).
\end{equation}
Assume $\scalebox{1.5}{$\chi$}_{H\ox}^{A} = \epsilon$. 
Then, from Eq.~(\ref{eq:relativechihox}), for all $\eta^E$ on $E$ and $\rho ^A$ on $A$,
\begin{equation}
\begin{split}
D(\mathcal{N_{\eta}}^{A \rightarrow B}(\rho ^A) || \mathcal{M}_{\omega_{\eta}}^{A \rightarrow B}(\rho^A)) & \leq  \epsilon .
\end{split}
\end{equation}
where $\omega_{\eta}^{B} := \mathcal{N_{\eta}}^{A \rightarrow B}(\rho_{avg,\eta} ^A)$ and $\rho_{avg,\eta}^{A}$ is the average of the optimal ensemble that achieves Holevo information for $\mathcal{N_{\eta}}$.
From the quantum Pinsker inequality~\cite{OP93}, for all $\rho ^A$ on $A$ 
\begin{equation}
\begin{split}
\left\| \mathcal{N_{\eta}}^{A \rightarrow B}(\rho ^A) - \mathcal{M}_{\omega_{\eta}}^{A \rightarrow B}(\rho ^A) \right\|_{1} & \leq  \sqrt{2 \epsilon \ln 2}.
\end{split}
\end{equation}
From Eq.~(\ref{eq:inducedtracenorm}), we have
\begin{equation}
\begin{split}
\left\| \mathcal{N_{\eta}}^{A \rightarrow B} - \mathcal{M}_{\omega_{\eta}}^{A \rightarrow B} \right\|_{1\rightarrow 1} & \leq  \sqrt{2 \epsilon \ln 2}.
\end{split}
\end{equation}
Using the relation between the induced trace norm and the diamond norm as expressed by Eq.~(\ref{eq:diamondtracenorm}), gives
\begin{equation}
\begin{split}
\left\| \mathcal{N_{\eta}}^{A \rightarrow B} - \mathcal{M}_{\omega_{\eta}}^{A \rightarrow B} \right\|_{\diamond} & \leq  \sqrt{2 \epsilon \ln 2}\cdot d.
\end{split}
\end{equation}
From the left half of the continuity bound in
 Lemma~\ref{lemma:continuity}, we have
\begin{equation}
\left\| (\1 ^{B} \ox U^{F}) W - \SWAP \right\| _{\rm op} \leq (2 \epsilon \ln 2)^{\frac{1}{4}} \sqrt{d},
\end{equation}
where the minimum is achieved by $U^{F}$.
 The channels $\mathcal{P}_{\alpha}^{E \rightarrow B} (\eta^E) := \tr_{F} (W(\alpha \ox \eta)W^{\dag})$, and $\id^{E \rightarrow B}$ have the dilations $W$ and $\SWAP$ respectively. From the right half of the continuity bound, we get
 \begin{equation}
 \left\| \mathcal{P}_{\alpha}^{E \rightarrow B}- \id^{E \rightarrow B} \ \right\|_{\diamond}  \leq 2(2\epsilon \ln 2)^{\frac{1}{4}} \sqrt{d} =: \Delta, \quad  \forall \alpha^{A}.
 \end{equation}
Thus, from the continuity of $\chi$~\cite{LS09,Winter15} (see Eq.~(\ref{eq:continuitycapacities}))
\begin{equation}
 \label{eq:continuitychi}
  \left|\chi(\mathcal{P}_{\alpha}) - \chi(\id) \right| \leq 2\Delta\log d + (2 + \Delta) H_2\left(\frac{\Delta}{2 +\Delta}\right),
 \end{equation}
 which gives us
 \begin{equation}
  \chi(\mathcal{P}_{\alpha}) \geq \max \left\{ (1- 2\Delta) \log d - (2 + \Delta) H_2\left(\frac{\Delta}{2 +\Delta}\right) ,0 \right\}.
\end{equation}
Using
\begin{equation}
H_2(\Delta) \leq 2\sqrt{\Delta(1-\Delta)},
\end{equation}
and $\scalebox{1.5}{$\chi$}_{A\ox}^{H} \geq  \chi(\mathcal{P}_{\alpha})$ (see Eq.~(\ref{eq:CHprodtens})), we have
\begin{equation}
\scalebox{1.5}{$\chi$}_{A\ox}^{H} \geq \max \left\{ \log d - 2\Delta \log d - \sqrt{8\Delta} , 0 \right\}.
\end{equation}
Then, rewriting the above inequality in terms of $\epsilon$, 
\begin{equation}
\label{eq:tightrelation}
\scalebox{1.5}{$\chi$}_{A\ox}^{H} \geq \max \Big\{ f(\epsilon), 0 \Big\},
\end{equation}
where 
\begin{equation}
f(\epsilon) := \log d -  (2^9 d^2 \epsilon \ln 2 )^{\frac{1}{4}} \log d - \left(2^{17} d^2 \epsilon \ln 2 \right)^{\frac{1}{8}}. 
\end{equation}
The function $f(\epsilon)$ is non-negative for $\epsilon \in [0,\epsilon_0]$ with
\begin{equation}
\epsilon_0 := \frac{1}{2^{13} d^2 \ln 2} \left( \frac{\sqrt{2 + 2( \log d )^2} - \sqrt{2}}{\log d} \right)^8.
\end{equation}
 Putting everything together for $\epsilon \in [0,\epsilon_0]$, we arrive at
  \begin{equation}
\scalebox{1.5}{$\chi$}_{H\ox}^{A}  + \scalebox{1.5}{$\chi$}_{A\ox}^{H} \geq  \min_{\epsilon > 0}\Big\{ \epsilon + f(\epsilon) \Big\}.
\end{equation}
As the function $\epsilon + f(\epsilon)$ is monotonically decreasing in the interval $\epsilon \in (0, \epsilon_0 ]$, the minimum is attained at $\epsilon_0$, thus we obtain
\begin{equation}
\label{eq:uncertainty1}
\scalebox{1.5}{$\chi$}_{H\ox}^{A}  + \scalebox{1.5}{$\chi$}_{A\ox}^{H} \geq \frac{1}{2^{13} d^2 \ln 2} \left( \frac{\sqrt{2 + 2( \log d )^2} - \sqrt{2}}{\log d} \right)^8,
\end{equation} 
which is non-zero.
\end{proof}

\bigskip

 \begin{remark}
   We present the uncertainty relation in the form of Eq.~(\ref{eq:uncertainty1}) motivated by the well-known entropic uncertainty relation~\cite{MU88}. Actually we get a tighter lower bound on $\scalebox{1.5}{$\chi$}_{A\ox}^{H}$ as a function of $\scalebox{1.5}{$\chi$}_{H\ox}^{A} := \epsilon$ from Eq.~(\ref{eq:tightrelation}) as can be seen in the Fig.~\ref{fig:comparision}.
  \begin{figure}
\centering
\begin{subfigure}
\centering
 \includegraphics[width= 0.45\linewidth]{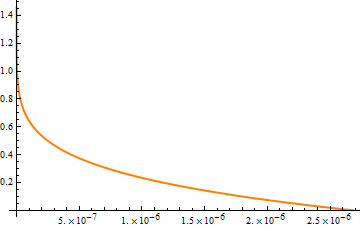}
  \label{fig:comparisiona}
\end{subfigure}
\quad
\begin{subfigure}
  \centering
  \includegraphics[width= 0.45\linewidth]{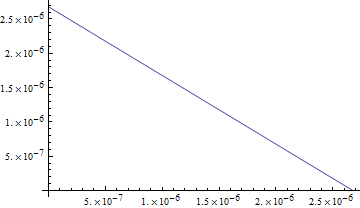}
  \label{fig:comparisionb}
\end{subfigure}
\caption{Plots of $\protect\scalebox{1.5}{$\chi$}_{A\ox}^{H}$ vs $\protect\scalebox{1.5}{$\chi$}_{H\ox}^{A}$ for dimension $d=3$. The left plot corresponds to Eq.~(\ref{eq:tightrelation}) and the right one to Eq.~(\ref{eq:uncertainty1}).}
\label{fig:comparision}
\end{figure}
\end{remark}

\bigskip

\begin{lemma}[Chen/Ji/Kribs/Zeng~\cite{CJK12}]\label{lemma:CJK}
Let $U$ be a random gate in $\rm U(d^2)$ according to the Haar measure, then for any $\delta > 0$,
\begin{equation}
\pr \left\{\vert S_{\rm min}(U) - \mathbb{E}(S_{\rm min}(U)) \vert \geq \delta\right\} \leq 2 \exp \left(- \frac{d^2 \delta^2}{64 (\log d )^2} \right),
\end{equation}
where $\mathbb{E}(S_{\rm min}(U))$ is the expectation value of the minimum output entropy.
Here (Corollary 44 of~\cite{CJK12}),
\begin{equation}
\mathbb{E}(S_{\rm min}(U)) \geq \log d - \frac{1}{\ln 2} - 1.
\end{equation}
\end{lemma}

\bigskip
  
\begin{remark}
When the $U$ are chosen according to the Haar measure on $\rm U(d^2)$, from Lemma~\ref{lemma:CJK} and Eq.~(\ref{eq:CHprodtensub}), we can give an upper bound on $\mathbb{E}(\scalebox{1.5}{$\chi$}_{H\ox}(U))$, the expectation value of the product state capacity with separable helper, which reads as follows
\begin{equation}
\begin{split}
\mathbb{E}(\scalebox{1.5}{$\chi$}^{A}_{H\ox}(U)) \leq 1 + \frac{1}{\ln 2},\\
\mathbb{E}(\scalebox{1.5}{$\chi$}^{H}_{A\ox}(U)) \leq 1 + \frac{1}{\ln 2}.\\
\end{split}
\end{equation}
It follows that when $d \rightarrow \infty$, by the concentration of measure phenomenon~\cite{HLW06}, with overwhelming probability 
\begin{equation}
\scalebox{1.5}{$\chi$}^{A}_{H\ox} , \scalebox{1.5}{$\chi$}^{H}_{A\ox} \leq 2.5.
\end{equation}
\end{remark}

\bigskip

\begin{remark}
\label{remark:zerocapacity}
The classical capacity of a quantum channel is zero iff the channel maps all inputs to a constant output,  i.e. the output of the channel is independent of the input. This helps us to identify the unitaries which have $C_{H \ox} = 0$. These unitaries must have effective channels with constant output for every choice of the initial environment state. At least in the case when $|A| = |F|$ and $|B| = |E|$, the unitary is the $\SWAP$.
Furthermore, for these unitaries, $C_{H}(\SWAP) = C_{H\ox}(\SWAP) = 0$.
\end{remark}


 \subsection{Controlled-unitaries}
 \label{sec:cu}

As we have noticed, the above defined passive assisted capacities, like the standard classical capacity of a quantum channel~\cite{Hastings09} (cf.~\cite{Wilde11}), admit multi-letter characterizations, thus posing a hard optimization problem. 
It is therefore important to single out classes of unitaries, if any, for which we can reduce to the single-letter case. 
We focus on controlled-unitaries, which apart from allowing for a simple characterization of capacities, provide examples for interesting  phenomena like super-additivity. 

\bigskip

\begin{definition}
We say that a unitary operator $U$ is \emph{universally entanglement-breaking} (resp. \emph{universally classical-quantum}), if for every $\ket{\eta} \in E$, 
  the effective channel $\mathcal{N}_{\eta}:\mathcal{L}(A) \rightarrow \mathcal{L}(B)$ 
  is entanglement-breaking (resp. classical-quantum).
  The set of universally entanglement-breaking  (resp. universally classical-quantum)  unitaries
  is denoted  $\mathfrak{E}$ (resp. $\mathfrak{CQ}$).
\end{definition}

\bigskip

For these unitaries, $C_{H\ox}$ reduces to the single-letter case. Indeed, for any $W \in \mathfrak{E}$, we have
\begin{equation}
C_{H\ox}(W) = \max_{\{ p(x) ,\rho_x \},\eta^{E}} I\bigl( X : B \bigr)_{\sigma},
\end{equation}
where the mutual information is evaluated with respect to the state $\sigma := \sum\limits_{x} p(x) \proj{x} \ox \mathcal{N_{\eta}}(\rho_x)$ and the maximization is over the ensemble $\{ p(x) ,\rho_x \}$.
This follows from additivity of Holevo information for entanglement-breaking channels,
\begin{equation}
\max_{\{p(x^n),\rho_{x^n}^{A^n} \}, \eta^{E^n}}
                    I\bigl( X^n : B^n \bigr)_{\sigma} = \sum\limits_i \max_{\{p(x),\rho_{x}^{A} \} , \eta_i} I \bigl( X : B \bigr)_{\sigma_i}
\end{equation}

Let us define a unitary operator $U_c : A\ox E \longrightarrow B \ox F$ with $\vert A \vert = \vert B \vert = \vert E \vert = \vert F \vert = d$ as follows
\begin{equation}
\label{eq:Uc}
U_c := \sum\limits_{i} \ket{i}^F \bra{i}^A \ox U_{i}^{E \rightarrow B}. 
\end{equation}
  Here $\{\ket{i} \} $ denotes an orthonormal basis of $A$ and $U_i \in \rm U(d)$.
 When the initial environment state is $\ket{\eta}$, the Kraus operators of the effective channel $\mathcal{N}_{\eta} : \mathcal{L}(A) \rightarrow \mathcal{L}(B)$ are given by $K_{i} = U_{i}\ket{\eta}\bra{i}$. Thus $\mathcal{N}_{\eta}$ is a classical-quantum channel for each choice of $\ket{\eta}$, and as consequence $U_c \in \mathfrak{CQ}$.
Hence 
\begin{equation}
\label{eq:controlled-ucapacity}
C_{H\otimes}(U_c) = \max_{p_{i},\eta} S\left(\sum\limits_{i} p_{i}U_{i}\proj{\eta} U_{i}^{\dag}\right). 
\end{equation}
For $U_c \ox V_c : A'E'AE \rightarrow B'F'BF$, the Kraus operators of the effective channel $\mathcal{N}_{\eta} : \mathcal{L}(A \ox A') \rightarrow \mathcal{L}(B \ox B')$, when the initial state of the environments $E'E$ 
is $\ket{\eta}$, are $K_{ij} = (U_{i}\otimes V_{j})\ket{\eta} \bra{ij}$ which is also a classical-quantum channel. Hence $U_c \ox V_c \in \mathfrak{CQ}$. We can also say
\begin{equation}
\label{eq:controlled-ucapacity2}
 C_{H\otimes}(U_c \ox V_c) = \max_{p_{ij}, \eta} S\left(\sum\limits_{i,j} p_{ij} (U_{i} \otimes V_{j})\proj{\eta} ( U_{i} \otimes V_{j})^{\dag}\right).
 \end{equation}

\bigskip
 
 \begin{remark}
Universal properties of bipartite unitary operators have been studied in~\cite{DNP15} although with different motivation than in this manuscript. As we are interested in evaluating environment-assisted capacities, we restrict the universal properties to pure environment states. We can extend the universal properties to a general density operators in the case of entanglement-breaking, and classical-quantum because of the convexity of these set of maps. In particular, they treat in full generality the question of bipartite unitaries which give constant channels for all input-environment states (see Theorem 2.4, Remark 2.5 of~\cite{DNP15}) (cf. Remark~\ref{remark:zerocapacity} in which we restricted to the case when $|A| = |F|$ and $|B| = |E|$).
 \end{remark}
 
 \bigskip

\begin{remark}
We have identified unitaries with $C_{H} = 0$. It is much harder to characterize unitaries with  quantum capacity $Q_{H} =0$~\cite{KMWY14}. $\SWAP$ was the only unitary which was identified  to have $Q_{H} = 0$. It was also conjectured there that $\sqrt{\SWAP}$ has zero passive environment assisted capacity. From the previous discussions $U_c$ has zero \emph{passive environment assisted} quantum capacity. When an arbitrary initial environment state $\ket{\eta}^{(n)}$ is input across $E^n$, the effective channel $\mathcal{N}_{\eta ^{(n)}} : A^n \rightarrow B^n$ is classical-quantum channel, thus the quantum capacity of the effective channel is $Q(\mathcal{N}_{\eta ^{(n)}}) = 0$. As a consequence of coding theorems for transmission of quantum information with a passive separable helper ($Q_{H\ox}$) and passive helper ($Q_{H}$), they are related by $Q_{H}(U_c) = \lim_{n \rightarrow \infty} \frac{1}{n} Q_{H\ox}(U_{c}^{\ox n})$. Thus $Q_{H}(U_c) = 0$. 

\end{remark}

\bigskip

\subsubsection{Two-qubit unitaries} 

 Here we evaluate the passive environment assisted classical capacity with separable 
helper for universally classical-quantum two-qubit unitaries.
A two-qubit controlled-unitary of the form $U_{c(2)} := \sum\limits_{i =0}^{1} \ket{i}^F \bra{i}^A \ox U_{i}^{E \rightarrow B} $, 
where  $U_i \in {\rm SU}(2)$.
Then, according to the parametrization reported in Appendix \ref{Upara},  the $U_{c(2)}$ has a parametric representation $\left( \frac{\pi}{2},\frac{\pi}{2},u \right)$
 where $0 \leq u \leq \frac{\pi}{2}$. 
From the previous discussion we know, $U_{c(2)} \in \mathfrak{CQ}$ when $0 \leq u \leq \frac{\pi}{2}$.
 Let the initial state of the environment be $\ket{\psi}^{E} =c_0 \ket{0} + c_1 \ket{1}$ with $c_0,c_1\in\mathbb{C}$ and $ \vert c_0 \vert ^{2} + \vert c_1 \vert ^{2} = 1$.
The input ensemble is 
 $\ket{0}, \ket{1}$ with probability $1- q, q$ respectively, then the capacity  is given by
\begin{equation}
\label{eq:cutwoqubit}
C_{H\otimes}(U_{c(2)}) = \max_{|c_0|^2,q} H_{2}\left(\frac{1}{2} + \sqrt{\frac{1}{4} - 4|c_0|^2|c_1|^2q(1- q)\cos^{2}u}\right).
\end{equation}
The maximum is attained at $|c_0|^2=q  = \frac{1}{2}$ and  it is
\begin{equation}
\label{eq:Uc2capacity}
C_{H\otimes}(U_{c(2)}) = H_{2}\left(\frac{1+ \sin u}{2} \right).
\end{equation}


\subsection{Super-additivity}
\label{subsec:super-additivity}

In this Subsection, we find two unitaries such that when they are used in conjunction, and their initial environments are entangled, they transmit more classical information than the sum of the classical information transferred by them individually. This phenomenon is called \emph{super-additivity}.

\begin{figure}[ht]
\centering
\begin{tikzpicture}[scale=0.35]
\draw[thick, green] (0,0) -- (1,0);
\draw [thick, green](1,0) -- (3,2) -- (7,2);
\draw [thick, green](1,0) -- (3,-2) -- (7,-2);
\draw [thick, red](3,4)--(7,4); \draw[thick,red] (3,-4) -- (7,-4);
\draw[ultra thick] (7,1) rectangle (11,5); \draw[ultra thick] (7,-5) rectangle (11,-1);
\draw[thick, blue] (11,4) -- (17,4);
\draw[thick, blue] (11,-4) -- (17,-4);
\draw [thick, purple](11,2) -- (14,2); \draw[thick, purple](11,-2) -- (14,-2);
\node[cloud, cloud puffs=15.7, cloud ignores aspect, minimum width=0.5cm, minimum height=2cm, align=center, draw, fill=purple!50] (cloud) at (14,0){} ; 
\draw[fill = blue!50] (17,0) ellipse (0.5 and 4.4);
\node[left] at (0,0){$H$};
\node[above] at (5,4){$A'$}; \node[below] at (5,-4){$A$};
\node[below] at (5,2){$E'$}; \node[above] at (5,-2){$E$};
\node[below] at (12.8,2){$F'$}; \node[above] at (12.8,-2){$F$};
\node[above] at (14,4){$B'$}; \node[below] at (14,-4){$B$};
\draw (9,3) node[font = \fontsize{13}{14}\sffamily\bfseries]{$\SWAP$};
\draw (9,-3) node[font = \fontsize{40}{42}\sffamily\bfseries]{$V_c$};
\end{tikzpicture}
\caption{The inputs controlled by Alice are  $A'$ and $A$. Helen controls $E'$ and $E$, Bob's systems are labelled as $B'$ and $B$. 
           The inaccessible output-environment systems are labelled as $F'$ and $F$. 
            Helen inputs 
          an entangled state in $E' E$.}
\label{fig:super}
\end{figure}
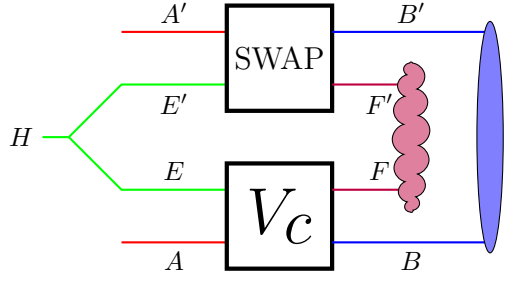

The following examples use the setting and notations  of Fig.~\ref{fig:super}.

\begin{enumerate}
\item Super-additivity for $C_{H\ox}$:

Let $V_c = \sum\limits_{i =0}^{2} \ket{i}^{F} \bra{i}^A \ox V^{E \rightarrow B}_{i}$ act on 2 qutrit systems and $V_{i}$  given as

\begin{equation}
V_0:=\left[\begin{array}{cccc}
1 & 0 & 0 \\
0 &  1 & 0 \\
0 & 0 & 1
\end{array}\right], \qquad
V_1:=\left[\begin{array}{cccc}
0 & 1 & 0 \\
1 &  0 & 0 \\
0 & 0 & 1
\end{array}\right], \qquad
V_2:=\left[\begin{array}{cccc}
1 & 0 & 0 \\
0 &  -1 & 0 \\
0 & 0 & 1
\end{array}\right].
\end{equation}
 We can use Eq.~(\ref{eq:controlled-ucapacity}) to  evaluate $C_{H\ox}(V_c)$, namely
\begin{equation}
C_{H\otimes}(V_c) = \max_{p_{i},\eta} S\left(\sum\limits_{i} p_{i}V_{i}\proj{\eta} V_{i}^{\dag}\right), 
\end{equation} 
where the maximization is over the initial state of the environment $\ket{\eta}^{E} = a \ket{0} + b \ket{1} + c\ket{2}$ with $\vert a \vert ^{2} + \vert b \vert ^{2} + \vert c \vert ^{2} = 1$ and $a,b,c \in \mathbb{C}$. Let $\ket{\psi_{i}} = V_{i} \ket{\eta}$. There are no $a,b,c \in \mathbb{C}$ such that the 
$\ket{\psi_{i}}$ are mutually orthogonal, thus $C_{H\ox}(V_c) < \log 3$.

Consider the scenario in Fig.~\ref{fig:super}, where Helen inputs a state $\ket{\Phi} = \frac{1}{\sqrt{2}} (\ket{00} + \ket{11})$ across $EE'$. In such a scenario the effective channel is $\mathcal{N}_{\Phi} : AA' \rightarrow BB'$ and when we input $\{ \ket{00}, \ket{01}, \ket{02} \}$ in $A'A$ the outputs in $B'B$ result respectively
\begin{equation}
\begin{split}
(\1\ox V_{0})\ket{\Phi} &= \frac{\vert 00 \rangle + \vert 11 \rangle}{\sqrt{2}},\\
 (\1 \ox V_{1})\ket{\Phi} &= \frac{\vert 01 \rangle + \vert 10 \rangle}{\sqrt{2}},\\
 (\1 \ox V_{2})\ket{\Phi} &= \frac{\vert 00 \rangle - \vert 11 \rangle}{\sqrt{2}},
\end{split}
\end{equation}
which are orthogonal, thus making the classical capacity of the effective channel equal to $\log 3$.
 Therefore 
 \begin{equation}
C_{H\otimes}(\SWAP \otimes V_c) > C_{H\ox}(\SWAP) + C_{H\ox}(V_c),
 \end{equation}
   since $\SWAP$ has zero passive environment assisted capacities (see Remark~\ref{remark:zerocapacity}).

\item Super-additivity for $C_{H}$: 

Let us consider $V_c: AE \rightarrow BF$ with $\vert A\vert = \vert F \vert = d^{2}, \vert E \vert =  \vert B \vert = d$, given by 
\begin{equation}
V_c = \sum\limits_{x,z} \vert xz \rangle^{F}\langle xz \vert^{A} \otimes (W(x,z))^{E \rightarrow B}.
\end{equation}
Here $W(x,z)$ are the discrete Weyl operators. From Eq.~(\ref{eq:controlled-ucapacity}) we have $C_{H\ox}(V_c) = \log d$. This is also the capacity with an unrestricted Helen as it saturates the dimension of $B$. Thus $C_{H}(V_c) = \log d$.
 Now consider $\SWAP \otimes V_c$ where $\SWAP : A'E' \rightarrow B'F'$ with $\vert A' \vert = \vert B' \vert = \vert E' \vert = \vert F' \vert = d$. 
 When Helen inputs a maximally entangled state $\ket{\Phi} = \frac{1}{\sqrt{d}} \sum\limits_{i=0}^{d-1} \ket{ii}$ across $E'E$, Alice inputs $\{ \ket{0ij} \}_{i,j=0}^{d-1}$ in $A' A$  (note that $|A|=d^2$), the outputs in $B'B$ are the set of states $\{\1 \ox X(i)Z(j) \ket{\Phi} \} $, which are $d^2$ orthonormal maximally entangled states in $B'B$.  Thus $C_{H}(\SWAP \otimes V_c) = 2\log d$ is achieved by the above inputs when they are chosen with equal probability of $\frac{1}{d^2}$. 
 In conclusion,
 \begin{equation}
 C_{H}(\SWAP \otimes V_c)) > C_{H}(V_c) + C_{H}(\SWAP).
\end{equation}  
\end{enumerate}

\section{Entanglement-environment-assisted capacity}
As we have noticed in the previous Section, $\SWAP$, in spite of having no communication capabilities with passive environment assistance on its own, can indeed enhance the classical communication when used in conjunction with other specific unitaries.  In other words $\SWAP$ acts like a ``dummy'' channel but helps to establish entanglement between the receiver and the initial environment,  as shown in Fig. \ref{fig:swap}.
  This is equivalent to sharing an entangled state between Helen and Bob which motivates us to rigorously define the following model of communication. 
\begin{figure}[ht]
\centering
\begin{tikzpicture}[scale=0.35]
\draw[thick, green] (0,0) -- (1,0);
\draw [thick, green](1,0) -- (3,2) -- (7,2);
\draw [thick, green](1,0) -- (3,-2) -- (7,-2);
 \draw[thick,red] (3,-4) -- (7,-4);
\draw[ultra thick] (7,-5) rectangle (11,-1);
\draw[thick, blue] (11,-4) -- (17,-4);\draw[thick,blue] (7,2) --(17,2);
 \draw[thick, purple](11,-2) -- (14,-2);
\node[cloud, cloud puffs=15.7, cloud ignores aspect, minimum width=0.3cm, minimum height=1cm, align=center, draw, fill=purple!50] (cloud) at (14,-2){} ; 
\draw[fill = blue!50] (17,-1) ellipse (0.5 and 3.4);
\node[left] at (0,0){$H$};
 \node[below] at (5,-4){$A$};
\node[below] at (5,2){$E'=K$}; \node[above] at (5,-2){$E$};
 \node[above] at (12.8,-2){$F$};
 \node[below] at (14,-4){$B$};\node[above] at (14,2){$B'=K$};
\draw (9,-3) node[font = \fontsize{40}{42}\sffamily\bfseries]{$V$};
\end{tikzpicture}
\caption{In Fig.~\ref{fig:super} when Helen inputs an entangled state across $E'E$ and an arbitrary state
  in $A'$, the $\SWAP$ acts like a ``dummy'' channel but helps to establish 
  entanglement between the receiver $BB'$ and the environment $E$.
  This is equivalent to sharing an entangled state between  Helen and Bob.}
\label{fig:swap}
\end{figure}
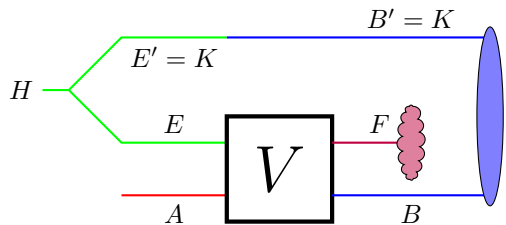

\begin{figure}[ht]
\begin{center}
\begin{tikzpicture}[scale=0.3]
\draw [ultra thick] (2,-5) rectangle (7,5) ;
\draw [ultra thick] (15,-8) rectangle (20,5) ;
\draw [ultra thick] (10,3) rectangle (12,5) node[midway]{$\mathcal{N}$};
\draw [ultra thick] (10,0) rectangle (12,2) node[midway]{$\mathcal{N}$};
\draw [ultra thick] (10,-5) rectangle (12,-3) node[midway]{$\mathcal{N}$};
\draw [ultra thick, red] (0,0)   -- (2,0);
\draw [ultra thick, blue] (20,-1.5)   -- (22,-1.5);
\draw [thick, red] (7,4.5) -- (10,4.5);
\draw [thick, red] (7,1.5) -- (10,1.5);
\draw [thick, red] (7,-3.5) -- (10,-3.5);
\draw [thick,blue] (12,4) -- (15,4);
\draw [thick, blue] (12,1) -- (15,1);
\draw [thick, blue] (12,-4) -- (15,-4);
\draw [thick, green] (8,-7) -- (8,3.5) -- (10,3.5);
\draw [thick, green] (8,-7) -- (8.2,0.5) -- (10,0.5);
\draw [thick, green] (8,-7) -- (8.4,-4.5) -- (10,-4.5);
\draw [thick, blue] (8,-7) -- (15,-7);
\node [below] at (11.5,-7){$K$};
\node[left] at (0,0){$M$};
\node[right] at (22,-1.5){$\tilde{M}$};
\node[above] at (7.5,4.5){$A$};\node[above] at (7.5,1.5){$A$};\node[above] at (7.5,-3.5){$A$};
\node[below] at (9.3,3.5){$E$};\node[below] at (9.3,0.5){$E$};\node[below] at (9.3,-4.5){$E$};\node[above] at (14,4){$B$};\node[above] at (14,1){$B$};\node[above] at (14,-4){$B$};
\node[below] at (8,-7){$\kappa$};
\draw (4.5,0) node[font = \fontsize{40}{42}\sffamily\bfseries]{$\mathcal{E}$};
\draw (17.5,-1.5) node[font = \fontsize{40}{42}\sffamily\bfseries]{$\mathcal{D}$};
\draw [thick,dotted] (7.5,-0.5) -- (7.5,-1.5);
\draw[thick,dotted] (9.3,-1.5) -- (9.3,-2.5);
\draw[thick,dotted] (11,-0.9) -- (11,-1.9);
\draw[thick,dotted] (14,-1) -- (14,-2);
\end{tikzpicture}
\end{center}
\caption{The general form of a protocol to transmit classical information when the helper 
         and the receiver pre-share entanglement; $\mathcal{E}$ and $\mathcal{D}$ are the encoding 
         and decoding maps respectively, $\kappa$ is the initial state of the environments and system $K$.}
\label{fig:entinfotask}
\end{figure}
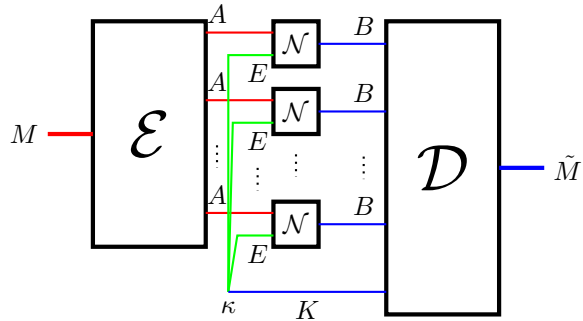

 By referring to Fig. \ref{fig:entinfotask}
an encoding CPTP map 
$\mathcal{E}: M \rightarrow \mathcal{L}(A^n)$ can be realised by preparing states $\{\alpha_m \}$ to be input across $A^n$ of $n$ instances of the channel. Let $\mathsf{M}$ denote the random variable corresponding to Alice's choice of message and $M$ be the associated Hilbert space with the orthonormal basis $\{ \ket{m} \}$.  A decoding CPTP map $\mathcal{D}:\mathcal{L}(B^n \ox K) \rightarrow \tilde{M}$ can be realised by a POVM $\{ \Lambda_{m} \}$. Here $\tilde{M}$ is Hilbert space associated to the random variable $\tilde{\mathsf{M}}$ for Bob's estimate of the message sent by Alice. The probability of error for a particular message $m$ is 
\begin{equation}
P_{e}(m) = 1 - {\tr}\left((\Lambda_{m})
({\cal N}^{\ox n} \ox {\id}_H)(\alpha_{m}^{A^n} \ox \kappa^{E^n K})\right).
\end{equation}

\bigskip

\begin{definition}
\label{def:entanglement-assisted-code}
 An \emph{entanglement-environment-assisted classical code} of block length $n$ is a family of triples
  $\{\alpha_{m}^{A^n},\kappa^{E^n K},\Lambda_{m}\}$ with error probability $\overline{P_e} :=\frac{1}{|M|}\sum_m P_e(m)$  and rate $\frac{1}{n}\log |M|$.
  A rate $R$ is achievable if there is a sequence of codes over their block length $n$ with $\overline{P_e}$ converging to $0$ and rate converging to $R$. The entanglement-assisted environment classical  capacity of $W$, denoted by $C_{EH}(W)$ or equivalently $C_{EH}(\mathcal{N})$, is the maximum achievable rate.
\end{definition}

\bigskip

\begin{theorem}
  \label{CEH}
  For an isometry $W:AE \longrightarrow BF$, the  entanglement-environment-assisted
  classical capacity is given by 
  \begin{equation}\begin{split}
    \label{eq:CEH}
    C_{EH}(W) &= \sup_{n} \max_{\kappa^{(n)}} \frac{1}{n} C(\mathcal{N}^{\ox n}_{\kappa^{(n)}}) \\
             &= \sup_{n} \max_{\{p(x),\alpha_{x}^{A^n}\},\kappa^{E^n K}}
                   \frac{1}{n} I\bigl( X : B^n K \bigr)_{\sigma},
  \end{split}\end{equation}
  where the mutual information is evaluated with respect to the state
  \begin{equation}
  \sigma = \sum\limits_{x} p(x) \proj{x} \ox \mathcal{N}^{\ox n}_{\kappa^{E^n K}}(\alpha^{A^n}_{x})
\end{equation}  
and the maximization is over the ensemble $\{ p(x) , \alpha^{A^n}_{x} \}$ and pure
  environment input states $\kappa^{(n)}$ on $E^n K$.
\end{theorem}

\medskip

\begin{proof}
The direct part (the ``$\geq$'' inequality), follows directly from the HSW Theorem~\cite{Holevo98,SW97} (cf.~\cite{Wilde11}).

 For the converse  part (the ``$\leq$'' inequality), consider a code of block length $n$ with  error probability $\overline{P_e}$. The state after encoding operation and action of the channel  is given by
\begin{equation}
\Phi^{M B^n K} = \frac{1}{|M|} \sum\limits_{m} p(m) \proj{m} \ox ( \mathcal{N}^{\ox n} \ox {\id}^K)(\alpha^{A^n}_{m} \ox \kappa^{E^n K}),
\end{equation}
and the state after decoding operation is given by
\begin{equation}
\omega^{M \tilde{M}} = \1^{M} \ox \mathcal{D}(\Phi^{M B^n K}).
\end{equation}
Then we have:
\begin{equation}
\begin{split}
 nR &= H(M)_{\omega}, \\
           &= I(M : \tilde{M})_{\omega} + H(M \vert\tilde{M})_{\omega}, \\
     &\leq I(M : \tilde{M})_{\omega} + H(\overline{P_e}) + nR\overline{P_e} ,\\
     &\leq I(M : B^n K)_{\Phi} + n\epsilon.
  \end{split}
\end{equation}

  The first inequality follows from  the application of Fano's inequality and the second one follows from the data processing inequality,
where $\epsilon  = \frac{1}{n} + R \overline{P_e}$. 
Setting $M = X$ we have
\begin{equation}
R \leq \frac{1}{n} I(X : B^n K) + \epsilon .
\end{equation}
 As $n \rightarrow \infty$ and $\epsilon \rightarrow 0$, the upper bound on the rate
  follows.
\end{proof}

 The classical capacity assisted by entangled states
of the form 
$\kappa^{E^n K^n} = \kappa^{E_1K_1}\ox\cdots\ox\kappa^{E_n K_n}$ in
Definition~\ref{def:entanglement-assisted-code}  is
denoted by $C_{EH\ox}(W)$, in analogy with $C_{H\ox}(W)$. Thus, we can say for $U_c$ (from Eq.~(\ref{eq:controlled-ucapacity2})),
\begin{equation}
C_{EH\ox}(U_c) = C_{H\ox}(\SWAP \ox U_c).
\end{equation}
As a consequence $C_{EH\ox}$ admits a single-letter characterization for $U_c$ given by
\begin{equation}
C_{EH\ox}(U_c) = \max_{p_{i}, \ket{\eta}} S\left(\sum\limits_{i} p_{i}(\1 \ox U_{i})\proj{\eta} (\1 \ox U_{i})^{\dag}\right).
\end{equation}

\bigskip

\begin{lemma}[D'Ariano/Lo Presti/Paris~\cite{DLP01}] \label{lemma:DLP}
 For two unitary operators $\{ U_1, U_2 \} \in \rm SU(2)$ with probability $\{ p_1, p_2 \}$, we have
\begin{equation}
\max_{\ket{\mu}} S\left(\sum\limits_{i} p_{i}U_{i}\proj{\mu} U_{i}^{\dag}\right) = \max_{\ket{\gamma}} S\left(\sum\limits_{i} p_{i}(\1 \ox U_{i})\proj{\gamma} (\1 \ox U_{i})^{\dag}\right),
\end{equation}
where $\ket{\mu}$ is a pure state in $A$ and $\ket{\gamma}$ is a  pure state in $AR$.
\end{lemma}

\bigskip

We can use Lemma~\ref{lemma:DLP} to evaluate $C_{EH \ox}$ for the universally classical-quantum two-qubit unitary interactions.
It results
\begin{equation}
\begin{split}
C_{EH\ox}(U_{c(2)}) &= C_{H\ox}(U_{c(2)}), \\
&= H_{2}\left(\frac{1+ \sin u}{2}\right),
\end{split}
\end{equation}
  which shows that entanglement does not enhance the classical capacity in this case. But clearly from the examples of super-additivity presented in the Subsection~\ref{subsec:super-additivity}, we can see that pre-shared entanglement  between Helen and Bob does indeed increase the classical communication capability.

\section{Conferencing sender and helper}
In this  Section we define the capacity with conferencing encoders, that is when Alice and Helen can freely communicate classical messages. A product state capacity with conferencing encoders is also defined when Alice and Helen are respectively restricted to product state encoding.

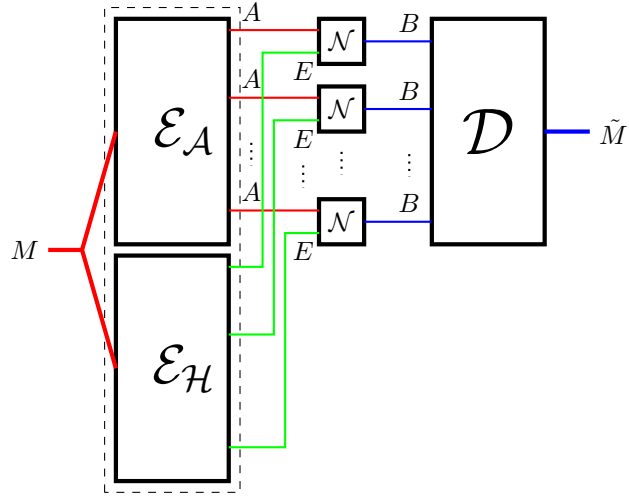
\begin{figure}[ht]
\begin{center}
\begin{tikzpicture}[scale=0.3]
\draw [dashed] (2.5,-16) rectangle (8.5,5.5) ;
\draw [ultra thick] (3,-5) rectangle (8,5) ;
\draw [ultra thick] (3,-15.5) rectangle (8,-5.5) ;
\draw [ultra thick] (17,-5) rectangle (22,5) ;
\draw [ultra thick] (12,3) rectangle (14,5) node[midway]{$\mathcal{N}$};
\draw [ultra thick] (12,0) rectangle (14,2) node[midway]{$\mathcal{N}$};
\draw [ultra thick] (12,-5) rectangle (14,-3) node[midway]{$\mathcal{N}$};
\draw [ultra thick, red] (0,-5.25)   -- (1.5,-5.25) -- (3,0);
\draw [ultra thick, red] (1.5,-5.25) -- (3,-10.5);
\draw [ultra thick, blue] (22,0)   -- (24,0);
\draw [thick, red] (8,4.5) -- (12,4.5);
\draw [thick, red] (8,1.5) -- (12,1.5);
\draw [thick, red] (8,-3.5) -- (12,-3.5);
\draw [thick,blue] (14,4) -- (17,4);
\draw [thick, blue] (14,1) -- (17,1);
\draw [thick, blue] (14,-4) -- (17,-4);
\draw [thick, green](8,-6) -- (9.5,-6) -- (9.5,3.5) -- (12,3.5);
\draw [thick, green] (8,-9) -- (10,-9) -- (10,0.5) -- (12,0.5);
\draw [thick, green](8,-14) --  (10.5,-14) -- (10.5,-4.5) -- (12,-4.5);

\node[left] at (0,-5.25){$M$};
\node[right] at (24,0){$\tilde{M}$};
\node[above] at (9,4.5){$A$};\node[above] at (9,1.5){$A$};\node[above] at (9,-3.5){$A$};
\node[below] at (11.3,3.5){$E$};\node[below] at (11.3,0.5){$E$};\node[below] at (11.3,-4.5){$E$};\node[above] at (16,4){$B$};\node[above] at (16,1){$B$};\node[above] at (16,-4){$B$};
\draw (6,0) node[font = \fontsize{22}{24}\sffamily\bfseries]{$\mathcal{E_{A}}$};
\draw (6,-10.5) node[font = \fontsize{22}{24}\sffamily\bfseries]{$\mathcal{E_{H}}$};
\draw (19.5,0) node[font = \fontsize{40}{42}\sffamily\bfseries]{$\mathcal{D}$};
\draw [thick,dotted] (9,-0.5) -- (9,-1.5);
\draw[thick,dotted] (11.3,-1.5) -- (11.3,-2.5);
\draw[thick,dotted] (13,-0.9) -- (13,-1.9);
\draw[thick,dotted] (16,-1) -- (16,-2);
\end{tikzpicture}
\end{center}
\caption{Schematic of a general protocol to transmit classical information with conferencing  encoders; 
$\mathcal{E_{A}}$ 
    and $\mathcal{E_{H}}$  are the encoding maps of Alice and Helen respectively. The decoding map is $\mathcal{D}$.}
\label{fig:confinfotask}
\end{figure}

 By referring to Fig.~\ref{fig:confinfotask},
an encoding CPTP map $\mathcal{E}: M \rightarrow \mathcal{L}(A^n) \ox \mathcal{L}(E^n)$ can be thought of two local encoding maps performed by Alice and Helen respectively and given by 
$\mathcal{E_{A}}: M \rightarrow \mathcal{L}(A^n)$, $\mathcal{E_{H}}: M \rightarrow \mathcal{L}(E^n)$. These can be realised by preparing pure product states $\{ \ket{\alpha_m}  \ox \ket{\eta_m}\}$ to be input across $A^n$ and $E^n$ of $n$ instances of the channel. A decoding CPTP map $\mathcal{D}:\mathcal{L}(B^n) \rightarrow \tilde{M}$ can be realised by a POVM $\{ \Lambda_{m} \}$.
The probability of error for a particular message $m$ is 
\begin{equation}
P_{e}(m) = 1- {\tr}\left( \Lambda_{m}
 \mathcal{N^{\ox}}(\alpha_{m}^{A^n} \ox \eta^{E^n}_{m})\right).
\end{equation}

\bigskip

\begin{definition}
 A \emph{classical code for conferencing encoders} of block length $n$ is a family of triples
  $(\ket{\alpha_m}^{A^n},\ket{\eta_m}^{E^n},\Lambda_{m})$ with the error probability $\overline{P_e} :=\frac{1}{|M|}\sum_m P_e(m)$
   and rate $\frac{1}{n}\log |M|$.
  A rate $R$ is achievable if there is a sequence of codes over their block length $n$ 
  with $\overline{P_e}$ converging to $0$ and rate converging to $R$. The classical capacity with conferencing encoders of $W$ denoted by $C_{\rm conf}(W)$ or equivalently $C_{\rm conf} (\mathcal{N})$ is the maximum achievable rate. If the sender and helper are restricted to fully separable states $\alpha^{A^n}_{m}$ and $\eta^{E^n}_{m}$ , i.e.~convex
  combinations of tensor products $\eta^{E^n}_{m} = (\eta_{1m}^{E_1} \ox \cdots \ox \eta_{nm}^{E_n})$, and $\alpha^{A^n}_{m} = (\alpha_{1m}^{A_1} \ox \cdots \ox \alpha_{nm}^{A_n})$ for all $m$
  the largest achievable rate is denoted $C_{\rm conf\ox} (W) = C_{\rm conf\ox} (\mathcal{N})$ and henceforth referred to as \emph{ classical capacity with product conferencing encoders}. 
\end{definition}

\bigskip

\begin{theorem}\label{thm:confcap}
 For an isometry $W:AE \longrightarrow BF$, the classical capacity of conferencing encoders model is given by
  \begin{equation}
  \label{eq:conf}
    C_{\rm conf}(W)  = \sup_{n} \max_{\{ p(x), \alpha^{A^n}_{x} \ox \eta_{x}^{E^n} \}} \frac{1}{n} I\bigl( X : B^n \bigr)_{\sigma},
  \end{equation}
  where the mutual information is evaluated with respect to the state
  \begin{equation}
  \sigma = \sum\limits_{x} p(x) \proj{x} \ox \mathcal{N}^{\ox n}(\alpha^{A^n}_{x} \ox \eta^{E^n}_{x})
\end{equation}  
 and the maximization is over the ensemble $\{ p(x), \alpha^{A^n}_{x} \ox \eta_{x}^{E^n} \}$.
  The classical capacity with product conferencing encoders is given by
  \begin{equation}
  \label{eq:conften}
    C_{\rm conf\ox} (W)  = \max_{\{ p(x), \alpha^{A}_{x} \ox \eta_{x}^{E}\}}  I\bigl( X : B \bigr)_{\sigma},
  \end{equation}
  where the mutual information is evaluated with respect to the state
  \begin{equation}
  \sigma = \sum\limits_{x} p(x) \proj{x} \ox \mathcal{N}(\alpha^{A}_{x} \ox \eta^{E}_{x})
\end{equation}   
and the maximization is over the ensemble $\{ p(x), \alpha^{A}_{x} \ox \eta_{x}^{E}\}$.
\end{theorem}

\medskip

\begin{proof}
 The direct part, ``$\geq$'' inequality, of the coding theorem follows from the HSW Theorem~\cite{Holevo98,SW97}(cf.~\cite{Wilde11}). 
  For the converse  part, ``$\leq$'' inequality, consider a code of block length $n$ with  error probability $\overline{P_e}$. The state after encoding operation and action of the channel is given by
 \begin{equation}
\Phi^{M B^n} = \frac{1}{|M|} \sum\limits_{m} p(m) \vert m \rangle^M\langle m \vert \ox \mathcal{N}^{\ox n}(\alpha^{A^n}_{m} \ox \eta^{E^n}_{m}),
\end{equation}
and the state after decoding operation is given by
\begin{equation}
\omega ^{M \tilde{M}} = \1^{M} \ox \mathcal{D}(\Phi ^{M B^n}).
\end{equation}
We then have:
\begin{equation}
\begin{split}
nR &= H(M)_{\omega}, \\
           &= I(M : \tilde{M})_{\omega} + H(M \vert\tilde{M})_{\omega}, \\
     &\leq I(M : \tilde{M})_{\omega} + H(\overline{P_e}) + nR\overline{P_e} ,\\
     &\leq I(M : B^n)_{\Phi} + n\epsilon.
\end{split}
\end{equation}
  The first inequality follows from the application of Fano's inequality and the second one follows from the data processing inequality, where $\epsilon  = \frac{1}{n} + R \overline{P_e}$.
   Setting $M = X$ we have
\begin{equation}
R \leq \frac{1}{n} I(X : B^n) + \epsilon .
\end{equation}
 As $n \rightarrow \infty$ and $\overline{P_e} \rightarrow 0$, the upper bound on the rate
  follows for $\conf$. For $\conften$ we have an additional  step, namely the additivity of mutual information.
\end{proof}

From Theorem \ref{thm:confcap} it is also clear that $C_{\rm conf} (W) = \lim_{n\rightarrow\infty} \frac1n C_{\rm conf\ox}(W^{\ox n})$.

\bigskip

\begin{remark}
In classical information theory, conferencing encoders for MAC were introduced in \cite{Willems83} where coding theorems were provided. Here each sender can gain partial knowledge of the other sender(s) message through conferencing, i.e.~a noiseless exchange of messages, eventually constrained to occur at a given rate. This is an example of ``cooperation" which is receiving an increasing attention in classical communication systems (see for e.g.~\cite{KMY06}), while it is still very rarely considered in the quantum domain. An exception is provided by~\cite{BN13} where the results of~\cite{Willems83} have been extended to classical-quantum MAC (both the inputs being classical and output quantum). 
In the conferencing encoders model, unlike~\cite{BN13}, we assume free classical communication between Alice and Helen, with both of them aiming to send the same message. We must remark here that the use of this resource, i.e.~free classical communication between Alice and Helen, does not trivialize the task as the global input state is still restricted to the set of separable states.
\end{remark}


\subsection{Role of entanglement in conferencing models (super-additivity)}

Entanglement played a peculiar role in the passive environment-assisted capacities and entanglement-environment-assisted capacities. We shall see in this Section that this is also true for the case of conferencing encoders. We consider the following example to highlight the role of entanglement with conferencing encoders.
 
Let us assume $\vert A \vert = \vert B \vert = \vert E \vert = \vert F \vert = d$.
From Eq.~(\ref{eq:conften}) we see that
\begin{equation}
\label{eq:conftenub}
C_{\rm conf\ox}(U) \leq \log d - S_{\rm min}(U),
\end{equation}
where $S_{\rm min}(U)$ is the minimum output entropy of $U$ as defined in Eq.~(\ref{eq:minetropy}).

\medskip
\begin{figure}[ht]
\begin{center}
\begin{tikzpicture}[scale=0.5]
\draw[thick, green] (0,-6) -- (3,-6); \draw[thick, red] (0,-4) -- (3,-4);
\draw [ultra thick] (3,-6.5) rectangle (5,-3.5) node[font = \fontsize{22}{24}\sffamily\bfseries][ midway] {$U$} ;
\draw[thick,purple] (5,-6) -- (13,-6);
\draw [thick,blue] (5,-4) -- (5.5,-4); \draw [thick,blue] (6.5,-4) -- (7.5,-4); \draw [thick,blue] (8.5,-4) -- (11,-4);
\draw[thick,purple] (0,0) -- (13,0); \draw[thick, purple] (0,-1) -- (13,-1);
\draw[ultra thick] (2,-8) rectangle (9,1);
\draw[ultra thick] (5.5,-4.5) rectangle (6.5,-3.5) node[midway] {$X$};
\draw[ultra thick] (7.5,-4.5) rectangle (8.5,-3.5) node[midway] {$Z$};
\draw[thick] (6,-1) -- (6,-3.5); 
\draw[thick] (8,0) -- (8,-3.5);
\draw (5.5,-8.5) node[font = \fontsize{22}{24}\sffamily\bfseries][below] {$U^{\rm aug}$} ;
\node [left] at (0,-6) {$E$};
\node [left] at (0,-4) {$A$};
\node [right] at (11,-4) {$B$};
\node [below] at (11,-6) {$F$};
\draw (0,-0.5) node[font = \fontsize{22}{24}\sffamily\bfseries][left] {$\{$} ;
\node [left] at (-0.75,-0.5) {$L$};
\node[cloud, cloud puffs=15.7, cloud ignores aspect, minimum width=0.6cm, minimum height=3.5cm, align=center, draw, fill=purple!50] (cloud) at (13,-3){} ; 
\end{tikzpicture}
\end{center}
  \caption{Shor's augmented unitary $U^{\rm aug}$ of a given unitary $U$ is depicted by the above quantum circuit. Here $X$ is the cyclic shift operator and $Z$, the phase operator as defined in Eq.~\ref{eq:cyclicphaseoperators}.}
  \label{fig:augunitary}
\end{figure}
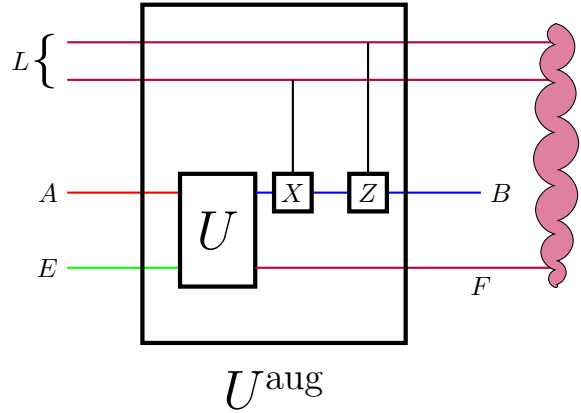

\bigskip

\begin{lemma}[Shor~\cite{Shor04}] For a given unitary $U: A\ox E \longrightarrow B \ox F$, the Shor's augmented unitary $U^{\rm aug} : A_0\ox E \longrightarrow B \ox F_0$ with $A_0 = L \ox A$ and $F_0 = F \ox L$, where $\vert L \vert = d^2$, is depicted by the quantum circuit in Fig.~\ref{fig:augunitary}.
Then for any environment state $\eta$, the effective channel for $U^{\rm aug}$ is given by 
\begin{equation}
\mathcal{N}^{\rm aug}_{\eta}(\sigma^{L} \ox \rho^{A}) := \sum\limits_{x,z}W(x,z)\mathcal{N}_{\eta}(\rho^{A})W(x,z)^{\dag} \bra{xz}  \sigma \ket{xz}.
\end{equation} 
 Here $W(x,z)$ are the discrete Weyl operators.
Then, for the augmented unitary, 
\begin{equation}
\label{eq:augconf}
C_{\rm conf\ox} (U^{\rm aug}) = \log d - S_{\rm min}(U).
\end{equation}
\end{lemma}

\bigskip

For \emph{any} given unitaries $W_1 : A \ox E \longrightarrow B \ox F$ and $W_2 : A' \ox E' \longrightarrow B' \ox F'$ we can ask whether the product conferencing encoders capacity is additive, i.e. whether the following equality holds true
\begin{equation}
C_{\rm conf\ox}(W_1 \ox W_2) \? C_{\rm conf\ox}(W_1) + C_{\rm conf\ox}(W_2).
\end{equation}
It is trivial to see that
\begin{equation}
C_{\rm conf\ox}(W_1 \ox W_2) \geq C_{\rm conf\ox}(W_1) + C_{\rm conf\ox}(W_2).
\end{equation}
Now with the following example we show that the above inequality is strict in general.

\bigskip

\begin{example}
 From Lemma~\ref{lemma:CJK}, we can guarantee the existence of a unitary $V : A \ox E \longrightarrow B \ox F$ (here all the parties are of equal dimension $d$) with the following lower bound on the minimum entropy:
 \begin{equation}
 S_{\rm min}(V) \geq \log d - \frac{1}{\ln 2} - 1.
 \end{equation}
 Now consider $V^* : A' \ox E' \longrightarrow B' \ox F'$  where the primed systems are isomorphic to the unprimed ones. Here $V^*$ is the conjugate of $V$. It is useful to note that 
 \begin{equation}
 S_{\rm min}(V) = S_{\rm min}(V^{*}).
 \end{equation}
 The unitaries of interest are the Shor augmented unitaries $V^{\rm aug}$ and $V^{* \rm aug}$. From Eq.~(\ref{eq:augconf}) we have
 \begin{equation}
 \begin{split}
 C_{\rm conf\ox}(V^{\rm aug}) &\leq \frac{1}{\ln 2} + 1, \\
 C_{\rm conf\ox}(V^{* \rm aug}) &\leq \frac{1}{\ln 2} + 1
 \end{split}
 \end{equation}
Let us evaluate the product conferencing encoders capacity of $V^{\rm aug} \ox V^{* \rm aug}$. As $V^{\rm aug} \ox V^{* \rm aug}$ is isomorphic to $(V \ox V^*)^{\rm aug}$, from Eq.~(\ref{eq:augconf}), we have
 \begin{equation}
 \label{eq:augconf1}
 \begin{split}
 C_{\rm conf\ox}(V^{\rm aug} \ox V^{* \rm aug}) &= C_{\rm conf\ox}((V \ox V^*)^{\rm aug}), \\
 &= 2\log d - S_{\rm min}(V \ox V^*).
 \end{split}
 \end{equation}
When Alice inputs a maximally entangled state across $AA'$, denoted by $\ket{\Phi}^{AA'}$, Helen inputs a maximally entangled state across $EE'$, denoted by $\ket{\Phi}^{EE'}$ the following holds true
\begin{equation}
\ket{\Phi}^{AA'} \ox \ket{\Phi}^{EE'} = \ket{\Phi}^{(AE)(A'E')}.
\end{equation}
Also
\begin{equation}
\begin{split}
V \ox V^* (\ket{\Phi}^{(AE)(A'E')}) &= \ket{\Phi}^{(AE)(A'E')}, \\
&= \ket{\Phi}^{AA'} \ox \ket{\Phi}^{EE'}.
\end{split}
\end{equation}
Thus $S_{\rm min}(V \ox V^*) = 0$, and from Eq.~(\ref{eq:augconf1}), we have
\begin{equation}
C_{\rm conf\ox}(V^{\rm aug} \ox V^{* \rm aug}) =  2 \log d,
\end{equation}
 exhibiting the role of entanglement in enhancing conferencing communication. We would like to emphasise that in this example entanglement enables us to send the entire bandwidth, without which we can only send paltry amount of information.
\end{example}


\subsection{ The classical capacity with conferencing encoders is always non-zero}

From Remark~\ref{remark:zerocapacity}, we have seen that $\SWAP$ has $C_{H} = 0$. Now, when conferencing is allowed, i.e. Alice and Helen are on the same footing as the sender of information, we can send classical information at the maximum rate. This motivates us to study whether some positive amount of classical information can always be transmitted with conferencing encoders.

From the definition of $C_{\rm conf\ox}$ and the previously defined quantities $\scalebox{1.5}{$\chi$} ^{A}_{H \ox}$,  $\scalebox{1.5}{$\chi$} ^{H}_{A\ox}$ (see Section~\ref{sec:model}) we can see that
\begin{equation}
C_{\rm conf\ox} \geq \max \{ \scalebox{1.5}{$\chi$} ^{A}_{H \ox},  \scalebox{1.5}{$\chi$} ^{H}_{A\ox} \}.
\end{equation}
Thus, for a unitary $U : A\ox E \longrightarrow B \ox F$ with $\vert A \vert = \vert B \vert = \vert E \vert = \vert F \vert =d$, we can invoke the uncertainty relation of Theorem~\ref{thm:uncertainty}, to give a lower bound on the $C_{\rm conf\ox}$ which reads as
\begin{equation}
C_{\rm conf\ox} \geq  \frac{\scalebox{1.5}{$\chi$}_{H\ox}^{A}  + \scalebox{1.5}{$\chi$}_{A\ox}^{H}}{2} \geq \frac{1}{2^{14} d^2 \ln 2} \left( \frac{\sqrt{2 + 2( \log d )^2} - \sqrt{2}}{\log d} \right)^8.
\end{equation}

 Now we derive a lower bound when the dimensions of $A,B,E,F$ are not equal.

\bigskip
  
  \begin{theorem}
Given a unitary $U :A \ox E \longrightarrow B \ox F$ with $\vert A \vert \vert E \vert  = \vert B \vert  \vert F \vert $, then 
\begin{equation}
C_{\rm conf\ox} \geq \frac{3}{8 \ln 2} \left(\frac{1}{|A| |E|}\right)^4.
\end{equation}
\end{theorem}
  
  \medskip

\begin{proof}
  Let $C_{\rm conf\ox} (U) = \delta$. Then, from  the quantum Pinsker inequality~\cite{OP93}, we have
  \begin{equation}
  \label{eq:normofproduct}
   \left\|\mathcal{N}(\alpha^A \ox \eta ^E)- \Omega^B \right\|_{1} \leq  \sqrt{2 \delta \ln 2}, \qquad \forall \alpha^A\otimes \eta^E,
  \end{equation} 
 where $\Omega^B := \sum p_{i} \mathcal{N}(\alpha_i^A \ox \eta_i ^E)$, the output of average of the ensemble $\{p_i, \alpha_i^A \ox \eta_i ^E \}$ which achieves the product conferencing capacity.
 Let us now consider the set of density operators $\{ \sigma^{Y}_{m.n} \}$ defined as follows:
 \begin{equation}
 \label{eq:spanoperators}
 \sigma^{Y}_{m,n} := 
 \begin{cases}
 \proj{m}, & \text{when } m =n \\
 \frac{1}{2}(\ket{m} + \ket{n})(\bra{m} + \bra{n}), & \text{when } m < n \\
 \frac{1}{2}(\ket{m} + i \ket{n})(\bra{m} - i \bra{n}), & \text{when } m > n
 \end{cases}
 \end{equation}
   Here $\{ \ket{m} \}$ denote the computational basis of the Hilbert space $Y$. The set 
 $\{ \sigma_{m,n} \}$ spans $\mathcal{L}(Y)$. Also $\{ \sigma^{A}_{m,n} \} \ox \{ \sigma^{E}_{o,p} \}$ spans $\mathcal{L}(A \ox E)$. Thus for an arbitrary state on $AE$, we can write
  \begin{equation}
\rho^{AE} = \sum\limits_{m,n = 1}^{\vert A \vert}\sum\limits_{o,p=1}^{\vert E \vert} \lambda_{m,n,o,p} \sigma^{A}_{m.n} \ox \sigma^{E}_{o,p}, 
\qquad
\sum\limits_{m,n = 1}^{\vert A \vert}\sum\limits_{o,p=1}^{\vert E \vert}  \lambda_{m,n,o,p} = 1.
\end{equation}
where
\begin{equation}
\label{eq:lambdaupperbound}
\vert \lambda_{m,n,o,p} \vert ^2 \leq \frac{1}{1 - \max \vert \tr(\sigma^{A}_{m,n} \sigma^{A}_{m',n'}) \tr(\sigma^{E}_{o,p} \sigma^{E}_{o',p'}) \vert ^2} = \frac{4}{3}.
\end{equation}
The maximization is over all the indices with at least one of the primed indices not equal to unprimed indices. From Eq.~(\ref{eq:spanoperators}) we can see that maximum is indeed reached for the case when exactly one primed index is different from the unprimed ones. Now
\begin{equation}  
\begin{split}
   \left\|\mathcal{N}(\rho^{AE})- \Omega^B \right\|_{1}  &= \left\| \sum\limits_{m,n = 1}^{\vert A \vert}\sum\limits_{o,p=1}^{\vert E \vert}  \lambda_{m,n,o,p} \mathcal{N}(\sigma^{A}_{m.n} \ox \sigma^{E}_{o,p}) - \sum\limits_{m,n = 1}^{\vert A \vert}\sum\limits_{o,p=1}^{\vert E \vert}  \lambda_{m,n,o,p}\Omega^B \right\|_{1}, \\
   &\leq \sum\limits_{m,n = 1}^{\vert A \vert}\sum\limits_{o,p=1}^{\vert E \vert}  \vert \lambda_{m,n,o,p} \vert \left\|  \mathcal{N}(\sigma^{A}_{m.n} \ox \sigma^{E}_{o,p}) - \Omega^B \right\|_{1}, 
   \end{split}
  \end{equation}
  which is due to the application of triangle inequality.
From Eq.~(\ref{eq:normofproduct}) and Eq.~(\ref{eq:lambdaupperbound}) we have
 \begin{equation}  
   \left\|\mathcal{N}(\rho^{AE})- \Omega^B \right\|_{1}  \leq  \sqrt{\frac{8}{3} \delta \ln 2}(|A| |E|)^2.   
  \end{equation}

 Let us further choose two states $\rho_{i}^{AE} := U^{\dag} (\omega_{i}^{B} \ox \kappa_{i}^{F}) U$ with the property that $\left\|\omega_1- \omega_2 \right\|_{1} = 2 $, i.e they are perfectly distinguishable states. We will hence have
  \begin{equation}
   \left\|\mathcal{N}(\rho_{i}^{AE})- \Omega^B \right\|_{1} \leq  \sqrt{\frac{8}{3} \delta \ln 2}(|A| |E|)^2, \qquad i=1,2,
  \end{equation}
  from which it  results
    \begin{equation}
   \left\|\mathcal{N}(\rho_{1}^{AE})- \mathcal{N}(\rho_{2}^{AE}) \right\|_{1} \leq  2\sqrt{\frac{8}{3} \delta \ln 2}(|A| |E|)^2. 
  \end{equation}
 Hence, it must be  $ \sqrt{\frac{8}{3} \delta \ln 2}(|A| |E|)^2 \geq 1 $, otherwise we have a contradiction. \\
 This leads to $\delta \geq \frac{3}{8 \ln 2} \left(\frac{1}{|A| |E|}\right)^4$.
\end{proof}

\bigskip

\begin{remark}
When the $U$ are chosen according to the Haar measure on $\rm U(d^2)$, from Lemma~\ref{lemma:CJK} and Eq.~(\ref{eq:conftenub}), we can give an upper bound on $\mathbb{E}(C_{\rm conf\ox} (U))$, the expectation value of the classical capacity with product conferencing encoders, which reads as
\begin{equation}
\mathbb{E}(C_{\rm conf\ox} (U)) \leq 1 + \frac{1}{\ln 2}.
\end{equation}
It follows that when $d \rightarrow \infty$, by the concentration of measure phenomenon~\cite{HLW06}, with overwhelming probability
\begin{equation}
C_{\rm conf\ox} (U) \leq 2.5.
\end{equation}
\end{remark}

\bigskip
  
For two-qubit unitaries a much tighter lower bound can be found which actually coincides with the upper bound, so giving the classical capacity with conferencing encoders.

\bigskip

\begin{theorem} 
In the qubit case, i.e. $|A|=|E|=|B|=|F|=2$, for any unitary $U : A\ox E \longrightarrow B \ox F$ we have
\begin{equation}
\label{eq:conftentwoqubit}
C_{\rm conf\ox} (U) = C_{\rm conf} (U) = 1.
\end{equation}
\end{theorem}

\medskip

\begin{proof} 
Let $U$ be a two-qubit unitary.  For its adjoint $U^\dag$ we have
\begin{equation}\begin{split}
\label{eq:product}
 U^{\dag}(\ket{\phi}^{B} \ox \ket{0} ^{F})  = \ket{\Phi_{0}}^{AE},\\
U^{\dag}(\ket{\phi}^{B} \ox \ket{1} ^{F})  = \ket{\Phi_{1}}^{AE},
\end{split}\end{equation}
where $\ket{\Phi_{i}}^{AE}$ are generically entangled across $AE$.

 Now, note that the subspace spanned by $\ket{\Phi_{0}}^{AE}$ and $\ket{\Phi_{1}}^{AE}$  contains at least one product state~\cite{STV98}. 
Say that
\begin{equation}
c_{o} \ket{\Phi_{0}}^{AE} + c_{1}\ket{\Phi_{1}}^{AE}
\end{equation} 
 is a product state in $AE$  with $c_{0}, c_{1} \in\mathbb{C}$. Thus, from Eq.~(\ref{eq:product}) 
 \begin{equation}
 U^{\dag}(\ket{\phi}^{B} \ox  (c_{0} \ket{0}^{F} + c_{1}\ket{1}^{F}))
 \end{equation}
  is a product state in $AE$.
For each choice of $\ket{\phi}^{B}$ we can find 
 \begin{equation}
 \ket{\psi}^{F} := c_{0} \ket{0}^{F} + c_{1}\ket{1}^{F}
 \end{equation}
  such that $U^{\dag}(\ket{\phi}^{B} \ox \ket{\psi}^{F})$
 is a product state in $AE$.  Let $\ket{\psi _{0}}^{F}$ and $\ket{\psi_{1}}^{F}$ be such states for the choices $\ket{0}^{B}$ and $\ket{1}^{B}$ respectively of $|\phi\rangle^B$.
 Hence, for a given $U$, we can find two input states which are product across $AE$,
 \begin{equation}
 \begin{split}
 & U^{\dag}(\ket{0}^{B} \ox \ket{ \psi_{0}}^{F}), \\
 & U^{\dag}(\ket{1}^{B} \ox \ket{\psi_{1}}^{F}),
 \end{split}
 \end{equation}
 such that we have two orthogonal output signals in system  $B$, thus achieving the capacity of 1 bit.
\end{proof}

\bigskip

\begin{remark}
The capacities $C_H$, $C_{H\ox}$, $C_{EH} $, $C_{\rm conf}$ and $C_{\rm conf}$ are continuous in the
  channel, with respect to the diamond (or completely bounded) norm. 
  Concretely, if $\left\|\mathcal{N}- \mathcal{M} \right\|_{\diamond} \leq \epsilon$, 
  then we have:
  \begin{equation}
  \label{eq:continuitycapacities}
  \begin{split}
  \bigl| C_{H\ox}(\mathcal{N})-C_{H\ox}(\mathcal{M}) \bigr| &\leq  2\epsilon\log |B| + (2 + \epsilon) H_2\left(\frac{\epsilon}{2 +\epsilon}\right), \\
   \bigl| C_{H}(\mathcal{N})-C_{H}(\mathcal{M}) \bigr|  &\leq  2\epsilon\log |B| + (2 + \epsilon) H_2\left(\frac{\epsilon}{2 +\epsilon}\right), \\
  \bigl| C_{EH}(\mathcal{N})-C_{EH}(\mathcal{M}) \bigr| &\leq  2\epsilon\log |B| + (2 + \epsilon) H_2\left(\frac{\epsilon}{2 +\epsilon}\right), \\
  \bigl| C_{\rm conf}(\mathcal{N})- C_{\rm conf} (\mathcal{M}) \bigr| &\leq  2\epsilon\log |B| + (2 + \epsilon) H_2\left(\frac{\epsilon}{2 +\epsilon}\right), \\
  \bigl| C_{\rm conf\ox} (\mathcal{N})- C_{\rm conf\ox} (\mathcal{M}) \bigr| &\leq  2\epsilon\log |B| + (2 + \epsilon) H_2\left(\frac{\epsilon}{2 +\epsilon}\right).
  \end{split}
  \end{equation}

   Since each of these capacities are expressed in terms of the quantum mutual information of a 
  classical-quantum state, and the optimization is over extra parameters due to the initial environment state, 
  the above results can be obtained following the same arguments as in~\cite{LS09}; cf.~\cite{KMWY14}. One distinction being the usage of the improved Alicki-Fannes continuity bound for conditional entropy~\cite{Winter15} compared to the original form of Alicki-Fannes~\cite{AF04} as used in~\cite{LS09}.
\end{remark}


\section{Conclusions}
  
We have laid the foundations of classical communication with environment assistance at the input. In such a model a benevolent helper is able to select the initial environment state of the channel, modelled as unitary interaction.They admit multi-letter formula, both for the unrestricted and separable helper, which are hard to  compute. These capacities are continuous like the unassisted ones, which are special case of our model. We further identified a class of unitaries which admit single-letter formula for the transmission of classical capacity with separable helper. Also, we have shown super-additivity  for both $C_{H\ox}$ and $C_{H}$. Due to the unique role $\SWAP$ plays in the examples of super-additivity, we considered  entanglement-environment-assisted capacities, where there is a pre-shared entanglement between the helper and receiver. 

The $U_c$ (as defined in Section~\ref{sec:cu}) constitute an interesting class of unitaries which are universally classical-quantum ($\in \mathfrak{CQ}$). In fact the $C_{H\ox}$ and $C_{EH\ox}$ admits single-letter characterization. The  capacity can be related to the problem of distinguishability of unitaries, when Holevo quantity is a measure of distinguishability. When we consider the distinguishability as mentioned in~\cite{Yang05} (i.e. with ancillary system), this is equal to $C_{EH\ox}(U_c)$. So, the additivity of these quantities can be related to the additivity of distinguishability  for unitary operations.

We have introduced a conferencing encoders model where the sender and  the helper are equipped with LOCC. Like the previous environment assisted models, they admit a regularized  formulae and are continuous. For a given unitary we can always transmit non-zero amount of classical information using a conferencing helper model. It would be interesting to find unitaries (if they exist) such that $C_{H}^{A}$ and $C_{A}^{H}$ are small but $C_{\rm conf}$ is large. At least in the case of unitaries where all the parties have equal dimensions, we can rule out such a possibility. This is due to the fact that  a small $C_{H}^{A}$ implies a large $C_{A}^{H}$ due to an uncertainty type relation, thus making $C_{\rm conf}$ large. Furthermore, we have evaluated the classical capacity for conferencing encoders for two-qubit unitaries, which turns out to be 1 bit. The computation of unrestricted helper capacities $C_{H}, C_{EH}, C_{\rm conf}$ is a major open problem.

Finally, it is worth noticing that if Helen exploits entanglement across channel uses we get memory effects 
on communication, hence the present study can shed further light on the subject of memory quantum channels~\cite{CGLM14}.


\section*{Acknowledgements}
SK thanks the Universitat Aut\`onoma de Barcelona for kind hospitality. AW's work is supported by the 
European Commission (STREP ``RAQUEL''), the European Research 
Council (Advanced Grant ``IRQUAT''), the Spanish MINECO (project FIS2013-40627-P), with the support of FEDER funds, 
as well as by​ ​the ​​Generalitat de Catalunya CIRIT, project 2014-SGR-966. DY's work was supported by the European Research Council (Advanced Grant ``IRQUAT") and the NSFC (Grant No. 11375165).


\appendices

\section{Parametrization of two-qubit unitaries}\label{Upara}

A general two-qubit unitary interaction can be described by $15$ real parameters. 
For the analysis of classical capacities under consideration we follow the arguments 
used in~\cite{KC01} to reduce the parameters to $3$ by the action of local 
unitaries with some further observations in~\cite{KMWY14}. According to the definition of capacities, the local unitaries on $A$, $B$, $E$ and $F$ 
do not affect the environment-assisted classical capacity, as they could be incorporated 
into the encoding and decoding maps, respectively, or can be reflected in a different choice
of environment state.

\bigskip

\begin{lemma}[Kraus/Cirac~\cite{KC01}]
  Any two-qubit unitary interaction  $V^{AE}$ is equivalent, up to local unitaries before and
  after the  $V^{AE}$, to one of the form 
  \[\begin{split}
    U^{AE} &= \sum_k e^{-i\lambda_{k}}\proj{\Phi_{k}} ,   \\
           &= \exp -\frac{i}{2}\bigl(\alpha_x \sigma_x\ox\sigma_x 
                        +\alpha_y \sigma_y\ox\sigma_y 
                        +\alpha_z \sigma_z\ox\sigma_z\bigr), \\
             &=: U(\alpha_x, \alpha_y, \alpha_z),           
  \end{split}\]
where $\sigma_x$, $\sigma_y$ and $\sigma_z$ are the Pauli operators and $\frac{\pi}{2} \geq \alpha_x \geq \alpha_y \geq |\alpha_z | \geq 0 $.  Furthermore the $\lambda_k$s are
  \begin{equation}
  \begin{split}
    \lambda_{1} &:= \frac{\alpha_{x} - \alpha_{y} + \alpha_{z}}{2}, \\
    \lambda_{2} &:= \frac{-\alpha_{x} + \alpha_{y} + \alpha_{z}}{2}, \\
    \lambda_{3} &:= \frac{-\alpha_{x} - \alpha_{y} - \alpha_{z}}{2}, \\
    \lambda_{4} &:= \frac{\alpha_{x} + \alpha_{y} - \alpha_{z}}{2}, 
 \end{split}
  \end{equation}
  and  the $\vert\Phi_k\rangle$ are the so-called ``magic basis'' vectors~\cite{HW97} 
  \begin{equation}
  \label{eq:magicbasis}
  \begin{split}
   \vert \Phi_{1} \rangle &:= \frac{\vert 00 \rangle + \vert 11 \rangle}{\sqrt{2}}, \\
    \vert \Phi_{2} \rangle &:= \frac{-i(\vert 00 \rangle - \vert 11 \rangle)}{\sqrt{2}}, \\
    \vert \Phi_{3} \rangle &:= \frac{\vert 01 \rangle - \vert 10 \rangle}{\sqrt{2}}, \\
    \vert \Phi_{4} \rangle &:= \frac{-i(\vert 01 \rangle + \vert 10 \rangle)}{\sqrt{2}}.
    \end{split}
  \end{equation}
  This is of course the familiar Bell basis, but with peculiar phases.
\end{lemma}

\bigskip

Hence the parameter space given by
\begin{equation}
\label{Parspacetotal}
  \mathfrak{T}_{total} = \left\{ (\alpha_x,\alpha_y,\alpha_z) : 
                         \frac{\pi}{2} \geq \alpha_x \geq \alpha_y \geq |\alpha_z | \geq 0 \right\},
\end{equation}
describes all two-qubit unitaries up to local basis choice .
This forms a tetrahedron with vertices $(0,0,0)$, $(\frac{\pi}{2},0,0)$,
$(\frac{\pi}{2},\frac{\pi}{2},-\frac{\pi}{2})$ and $(\frac{\pi}{2},\frac{\pi}{2},\frac{\pi}{2})$.

As we are interested in evaluating the capacities of unitaries, we use,
\begin{equation}
\label{eq:complexconjugate}
  U\left(\alpha_{x},\alpha_{y},\frac{\pi}{2}\!+\!\alpha_{z}\right) 
    \!=\! 
  -i(\sigma_z \ox \1)\, U^*\left(\alpha_{x},\alpha_{y},\frac{\pi}{2}\!-\!\alpha_{z}\right) (\1 \ox \sigma_z),
\end{equation} 
where $U^*$ is the complex conjugate of $U$. 
Note that the latter has the same 
environment-assisted classical capacities; indeed, any code for $U$ is transformed
into one for $U^*$ by taking complex conjugates.
The reduced parameter  space given by
\begin{equation}
\label{Parspace}
  \mathfrak{T} = \left\{ (\alpha_x,\alpha_y,\alpha_z) : 
                         \frac{\pi}{2} \geq \alpha_x \geq \alpha_y \geq \alpha_z \geq 0 \right\},
\end{equation}
describes all two-qubit unitaries up to local basis choice and
complex conjugation (we should note that in general $U \ox V$ and $U \ox V^*$ have different environment assisted capacities and in such cases we should consider $\mathfrak{T}_{total}$, say for example to provide a complete characterization of super-additivity).
This forms a tetrahedron with vertices $(0,0,0)$, $(\frac{\pi}{2},0,0)$,
$(\frac{\pi}{2},\frac{\pi}{2},0)$ and $(\frac{\pi}{2},\frac{\pi}{2},\frac{\pi}{2})$.

Familiar two-qubit gates can easily be identified within this parameter space: 
for instance, $(0,0,0)$ represents the identity $\1$, 
$(\frac{\pi}{2},0,0)$ the CNOT, $(\frac{\pi}{2},\frac{\pi}{2},0)$ the DCNOT (double
controlled-not), and $(\frac{\pi}{2},\frac{\pi}{2},\frac{\pi}{2})$ 
the $\SWAP$ gate, respectively.

Consider the unitaries $U^{'}_c(d)$
 with $\alpha_x = 0, \alpha_y = 0, \alpha_z = d$ where $0 \leq d \leq \frac{\pi}{2}$. These unitaries lie outside the tetrahedron $\mathfrak{T}$. When expressed in matrix form, these unitaries are diagonal in the magic basis (same order as in Eq.~(\ref{eq:magicbasis})) with the diagonal elements $\{ e^{-i\frac{d}{2}}, e^{-i\frac{d}{2}}, e^{i\frac{d}{2}}, e^{i\frac{d}{2}} \}$. To see their parametric representation in the tetrahedron $\mathfrak{T}$, we follow the argument used in Appendix A of~\cite{HVC02} (cf. Example 12 of~\cite{KMWY14}).
 Observe that the spectrum of $U_c^{'T}U^{'}_c$ is 
$(e^{-2i\lambda_1} ,e^{-2i\lambda_2},e^{-2i\lambda_3},e^{-2i\lambda_4} )$, 
where the transpose operator is with respect to the magic basis. The spectrum of $U_c^{'T}U^{'}_c$ is thus $\left( e^{id},e^{id},e^{-id},e^{-id} \right)$. Using the order property 
$\frac{\pi}{2} \geq\lambda_4 \geq \lambda_1 \geq \lambda_2 \geq \lambda_3 \geq -\frac{3\pi}{4}$ 
(condition \eqref{Parspace} written in terms of $\lambda_k$) and solving the linear 
equations in $\alpha_x$, $\alpha_y$ and $\alpha_z$, we get the parametric points 
as $(d,0,0)$ which correspond to the edge joining the identity $\1$ and CNOT i.e., these unitaries are controlled-unitaries of the form $\sum\limits_{i=0}^{1} \ket{i}^B \bra{i}^A \ox U_{i}^{E \rightarrow F}$, 
where  $U_i \in {\rm SU}(2)$. Here the parameter $d$ is given by $d = t$ when $t \leq \frac{\pi}{2}$ and $d = \pi - t$ when $t \geq \frac{\pi}{2}$ where $2\cos t = \tr U_0^{\dag}U_1$.

Now when we apply $\SWAP$ to $U^{'}_c(d)$ i.e., the unitary of interest, $\SWAP \cdot U'_c(d) = U\left(\frac{\pi}{2}, \frac{\pi}{2}, \frac{\pi}{2} + d\right)$ is outside the parameter tetrahedron $\mathfrak{T}$. From Eq.~(\ref{eq:complexconjugate}), we get $U\left(\frac{\pi}{2}, \frac{\pi}{2}, \frac{\pi}{2} + d\right) = U^{*}\left(\frac{\pi}{2}, \frac{\pi}{2},\frac{\pi}{2}-d\right)$, upto local unitaries, which lie in the parameter space. In essence, up to local unitaries and complex conjugation $U_{c(2)} := \sum\limits_{i=0}^{1} \ket{i}^F \bra{i}^A \ox U_{i}^{E \rightarrow B} $ has parameters $(\frac{\pi}{2}, \frac{\pi}{2},u)$ where $u = \frac{\pi}{2} - d$.



\end{document}